\documentclass[12pt,draftcls,onecolumn,twoside]{IEEEtran}
\usepackage[T1]{fontenc}
\usepackage{ifpdf}
\usepackage{color}
\usepackage{psfrag}
\usepackage{nicefrac}
\usepackage{tikz}
\usepackage{pgfplots}
\pgfplotsset{compat=newest}
\usepackage{soul}			
\usepackage{enumerate}

\usepackage[noadjust]{cite}
\usepackage[cmex10]{amsmath}
\usepackage{amsfonts}
\usepackage{amssymb}
\usepackage{dsfont}    
\usepackage{mathtools} 
\usepackage{fixmath}  

\usepackage{algorithmic}

\usepackage{array}

\newcommand{\ve}[1]{{\boldsymbol{#1}}}
\newcommand{\ma}[1]{{\boldsymbol{#1}}}
\newcommand{\SNR}{\mathop{\mathrm{SNR}}\nolimits}
\newcommand{\Real}{\mathop{\mathrm{Re}}\nolimits}
\newcommand{\Imag}{\mathop{\mathrm{Im}}\nolimits}
\newcommand{\Prob}{{\mathbb{P}}}
\newcommand{\E}{{\mathbb{E}}}
\newcommand{\Cov}{\mathop{\mathbb{C}\mathrm{ov}}\nolimits}
\newcommand{\Var}{\mathop{\mathbb{V}\mathrm{ar}}\nolimits}

\newcommand{\erf}{\mathop{\mathrm{erf}}\nolimits}
\newcommand{\erfc}{\mathop{\mathrm{erfc}}\nolimits}
\newcommand{\e}{\mathop{\mathrm{e}}\nolimits}

\newtheorem{definition}{Definition}

\newtheorem{lemma}{Lemma}
\newtheorem{corollary}{Corollary}
\newtheorem{proposition}{Proposition}

\newtheorem{remark}{Remark}
\newtheorem{example}{Example}

\usepackage[tight,footnotesize]{subfigure}

\begin{document}
%
%
%
\title{Robust Analog Function Computation via Wireless Multiple-Access Channels}
%
\author{Mario~Goldenbaum,~\IEEEmembership{Student~Member,~IEEE,} and S\l
  awomir~Sta\'{n}czak,~\IEEEmembership{Senior Member,~IEEE}
\thanks{The authors are listed in alphabetical order. Parts of the material in this paper were presented at the IEEE
  Wireless Communications and Networking Conference, Budapest, Hungary, April
  2009 and the 43rd Asilomar Conference on Signals, Systems and Computers,
  Pacific Grove, CA, USA, November 2009.}%
\thanks{The authors are with the Fachgebiet für Informationstheorie und theoretische Informationstechnik, Technische Universität Berlin, and with the Fraunhofer Institute for Telecommunications Heinrich Hertz Institute, Berlin, Germany (e-mail: mario.goldenbaum@tu-berlin.de, slawomir.stanczak@hhi.fraunhofer.de).}}%
\IEEEpubid{}
\markboth{Submitted to IEEE TRANSACTIONS ON COMMUNICATIONS}%
{GOLDENBAUM AND STA\'{N}CZAK: ROBUST ANALOG FUNCTION COMPUTATION VIA MULTIPLE-ACCESS CHANNELS} 
%
\IEEEaftertitletext{\vspace{-1\baselineskip}}
\maketitle
%
%
%
%
%
%
%
\begin{abstract}
  Various wireless sensor network applications involve the computation of a
  pre-defined function of the measurements without the need for reconstructing
  each individual sensor reading. Widely-considered examples of such functions
  include the arithmetic mean and the maximum value. Standard
  approaches to the computation problem separate computation from communication: quantized sensor readings are
  transmitted interference-free to a fusion center that reconstructs each sensor reading
  and subsequently computes the sought function value. Such separation-based
  computation schemes are generally highly inefficient as a complete reconstruction of individual sensor readings is not necessary for the fusion center to compute a function of them. In particular, if the mathematical structure of the wireless
  channel is suitably matched (in some sense) to the function, then channel
  collisions induced by concurrent transmissions of different nodes can
  be beneficially exploited for computation purposes. Therefore, in this paper a practically relevant analog computation scheme is proposed that allows for an efficient estimate of linear and nonlinear functions over the wireless
  multiple-access channel. After analyzing the asymptotic properties of the estimation
  error, numerical simulations are presented to show the potential for huge
  performance gains when compared with time-division multiple-access based computation schemes.
\end{abstract}
%
%
%
\section{Introduction} \label{sec:introduction}
\IEEEPARstart{I}{n contrast} to traditional wireless networks, wireless sensor
networks are deployed to perform various application tasks such as
environmental monitoring or disaster alarm. Indeed, rather than transmitting
and reconstructing the data of each individual sensor node, wireless sensor
network applications often involve the computation of some pre-defined
function of these data (called sensor readings), which includes the arithmetic
mean, the maximum or minimum value, and different polynomials
\cite{Giridhar:Kumar:06}. In this paper, we address the problem of computing
functions over a wireless Multiple-Access Channel (MAC) with a fixed number of
sensor nodes and a single receiver that is referred to as the fusion center. A
standard approach to this computational problem widely used in contemporary
sensor networks is to let each sensor node transmit \emph{separately} a
quantized version of its sensor reading to the fusion center as a stream of
information-bearing symbols. The data rate at which each sensor node transmits
is chosen such that the fusion center can reconstruct each (quantized) sensor
reading perfectly and \emph{subsequently} computes the sought function. The
data transmission and the function computation are therefore completely
disjoint processes. Moreover, in order to perfectly reconstruct each sensor
reading, orthogonal medium access protocols such as Time-Division Multiple
Access (TDMA) are typically used for the data transmission to establish
interference-free connections between each sensor node and the fusion center
by avoiding the interference from other transmissions.

Separation-based medium access protocols are in general highly suboptimal when for
instance maximizing computation throughput defined as the rate at which
quantized sensor readings are reconstructed at the fusion center subject to
some communication constraints. In particular, the information-theoretic
result of \cite{Nazer:Gastpar:07b} suggest that the superposition property of
the wireless channel can be beneficially exploited if the MAC is
\emph{matched} in some mathematical sense to a function being computed. The
approach, which is known as \emph{Computation over MAC (CoMAC)}, can be seen
as a method for merging the processes of data transmission and function
computation by exploiting channel collisions induced by a concurrent access of
different nodes to a common channel. An immediate consequence of this approach
is a higher computation throughput, and with it a reduced latency or lower
bandwidth requirements.

The analysis in \cite{Nazer:Gastpar:07b} also shows that in CoMAC scenarios,
codes with a certain algebraic structure may outperform random codes. One such
an example can be found in \cite{Koerner:Marton:79} (see also
\cite{Nazer:Gastpar:07b}) where a receiver aims at decoding the parity of two
dependent binary messages. The code design is in this case driven by an
application which is the modulo-two sum computation, and therefore the example
lifts a strict separation between computation and communication. The research
on structured codes is however still in its infancy, with some work on codes
for computing functions that naturally match the mathematical structure of the
underlying channel. Note that due to the superposition property of the
wireless channel, the wireless MAC can be seen as a summation-type linear
operator mapping the input space to the set of complex-valued numbers. Hence
functions naturally matched to this channel are linear functions that
constitute only one class of functions of interest in practice.

In light of practical constraints, a serious drawback of the
information-theoretic approach in \cite{Nazer:Gastpar:07b} and other related
results (see also Section \ref{sec:related_work}) is the implicit assumption
that if two symbols are put on the channel input, then the corresponding
decoder observes the sum of these inputs. Obviously, this is satisfied in
additive white Gaussian channels with users perfectly synchronized on the
symbol and phase level. In practical wireless sensor networks, however, it may
be extremely difficult and expensive in terms of resources to ensure such a
perfect synchronization. Hence, even if structured codes were available, the
question remained how to exploit the superposition property of the wireless
channel in the presence of practical impairments.

In this paper, we propose and analyze a novel CoMAC scheme for wireless sensor
applications that requires only a \emph{coarse block-synchronization}, and
therefore it is robust against synchronization errors. It is a a simple analog
joint source-channel computation scheme, in which
\begin{enumerate}
\item each sensor node encodes its message (sensor reading) in the power of a
  series of random signal pulses, and
\item the receiver estimates the function value from the received power. 
\end{enumerate}
Another crucial advantage of the proposed analog computation scheme is its
ability to reliably and efficiently estimate non-linear functions of sensor
readings. We achieve such \emph{non-linear computational capabilities} by
letting each sensor node pre-process its sensor readings prior to
transmission, followed by a receiver-side post-processing of the received
signal, which is a noise-corrupted weighted sum of the pre-processed sensor
readings from different sensor nodes. The pre-processing functions and the
post-processing function are to be chosen so as to match the wireless channel
with its superposition property to a function that we intend to evaluate at
the sensor readings. The weights are due to the impact of the fading channel,
which needs to be compensated in practical systems. %
%
%
%
%
%
\subsection{Related Work} \label{sec:related_work}
%
%
In the context of sensor networks, viewed as a collection of distributed
computation devices, Giridhar and Kumar took the first steps towards a
theory-based framework for \emph{in-network computation} with the aim of
characterizing efficient application-specific computation strategies \cite{Giridhar:Kumar:06}. The work is however
focused on complexity and protocol aspects and does not explicitly take into
account the properties of wireless communication channels. A similar holds
true for \cite{Ying:Srikant:Dullerud:07},
\cite{Subramanian:Gupta:Shakkottai:07}, the information theoretical considerations in \cite{Orlitsky:Roche:01} and 
\cite{Doshi:Shah:Medard:Effros:10} as well as for \cite{Appuswamy:Franceschetti:Karamchandani:Zeger:11}, which is mainly devoted to wired
networks. In contrast, harnessing the explicit structure of the channel for reliable function computations was first thoroughly analyzed in \cite{Nazer:Gastpar:07b} with an emphasis on information theoretical insights, whereas computations over noiseless linear channels are considered in \cite{Keller:Karamchandani:Fragouli:10}. 

Function computation in sensor networks is a fundamental building block of
gossip and consensus algorithms, a form of distributed in-network data
processing aiming at achieving some network-wide objectives based on local
computations. Such algorithms, which compute a global function of sensor
readings and distribute the function values among the nodes, have attracted a
great deal of attention (see \cite{Boyd:Ghosh:Prabhakar:Shah:06}--\nocite{Olfati:Fax:Murray:07}\cite{Dimakis:Kar:Moura:Rabbat:Scaglione:10}
and references therein). Most gossip and consensus protocols, however, require interference free transmissions between adjacent nodes, except for the recent work in \cite{Kirti:Scaglione:Thomas:07}--\nocite{Nazer:Dimakis:Gastpar:11}\nocite{Nokleby:Bajwa:Calderbank:Aazhang:11}\cite{Goldenbaum:Boche:Stanczak:12c}, where it was shown that the superposition property of the wireless channel can be
advantageously exploited to accelerate convergence speeds.

In \cite{Gastpar:Vetterli:03}, an analog
joint source-channel communication scheme was proposed to exploit the
superposition property of the Gaussian MAC for the optimal estimation of some
desired parameter from a collection of noisy sensor readings. The approach
outperforms comparable digital approaches based on the standard separation
design principle between source and channel coding, as proposed by Shannon in
his landmark paper \cite{Shannon:48}. Extensions of the analog joint source-channel scheme to more general estimation problems in wireless networks can be found in
\cite{Mergen:Tong:06}--\nocite{Bajwa:Haupt:Sayeed:Nowak:07}\nocite{Stanczak:Wiczanowksi:Boche:07}\cite{Banavar:Tepedelenlioglu:Spanias:12},
whereas References
\cite{Li:Dai:07}--\nocite{Mergen:Naware:Tong:07}\nocite{Li:Evans:12}\cite{Banavar:Smith:Tepedelenlioglu:Spanias:12}
are devoted to the detection counterparts.

Finally, we point out that the basic idea of \emph{physical-layer network
  coding} is to exploit the superposition property of the wireless channel as
well. Indeed, in contrast to the traditional network coding principle applied
across the packets on the network layer, the physical-layer network coding
generates linear codewords immediately on the wireless channel by
superimposing electromagnetic waves from different, concurrently transmitting
and perfectly synchronized nodes \cite{Zhang:Liew:Lam:06b}--\nocite{Katti:Gollakota:Katabi:07}\cite{Nazer:Gastpar:11}.
%
%
%
%
\subsection{Paper Organization} \label{sec:organization}
%
Section \ref{sec:problem_statement} introduces the system model, formulates
the problem and provides definitions used in this paper. In Section
\ref{sec:analog_over}, we present a novel analog CoMAC scheme for estimating
linear and non-linear functions of sensor readings and study the estimation
error under the proposed scheme in Section \ref{sec:error_analysis}. This
analysis is used to define appropriate estimators for two function examples of
great practical importance: the arithmetic mean and the geometric
mean. Numerical examples in Section \ref{sec:num_examples} illustrate the
performance of the proposed CoMAC scheme and compares it with a TDMA-based
computation scheme to show the potential for huge performance gains under
different network parameters. Finally, Section \ref{sec:discussion} concludes
the paper.\footnote{\emph{Notation:} Random variables are denoted with
  uppercase letters, random vectors by bold uppercase letters, realizations by
  lowercase letters and vector valued realizations by bold lowercase letters,
  respectively. The sets of natural, nonnegative integer, real, nonnegative
  real, positive real, and complex numbers are denoted by $\mathds{N}$,
  $\mathds{Z}_+$, $\mathds{R}$, $\mathds{R}_+$, $\mathds{R}_{++}$,
  $\mathds{C}$. The distributions of normally distributed real and proper
  complex random elements are denoted by
  $\mathcal{N}_{\mathds{R}}(\cdot,\cdot)$ and
  $\mathcal{N}_{\mathds{C}}(\cdot,\cdot)$. $\mathcal{LN}(\cdot,\cdot)$
  denotes the log-normal distribution and $\chi^2_n$ the Chi-square
  distribution with $n$ degrees of freedom, respectively. The error function
  and error function compliment are described by $\erf(\cdot)$ and
  $\erfc(\cdot)$. $\mathds{1}_{\mathcal{B}}(x)$ denotes the indicator function
  on set $\mathcal{B}$. The imaginary unit is denoted by $i$ and hence
  $i^2=-1$.}
%
%
%
%
\section{Definitions, System Model and Problem
  Statement} \label{sec:problem_statement}
Throughout the paper, all random elements are defined over an appropriate
probability space $(\Omega,\mathcal{A},\Prob)$, with sample space $\Omega$,
$\sigma$-Algebra $\mathcal{A}$ of subsets of $\Omega$ and probability measure
$\Prob$ on $\mathcal{A}$. It is assumed that all functions of random variables
and stochastic processes are Borel functions to ensure that all resulting
random elements are well defined.

We consider a wireless sensor network consisting of $K\in\mathds{N}$ spatially
distributed single-antenna sensor nodes and one designated single-antenna
Fusion Center (FC)
. Without loss of generality (w.l.o.g.) it is assumed that the $K$ nodes are
identical and we use $\mathcal{K}\coloneqq\{1,\dots,K\}$ to denote the set of
all sensor nodes (numbered in an arbitrary order). Basically the sensor nodes have the task to jointly observe a certain physical
phenomenon (e.g., temperature, pressure, humidity, acceleration, illumination)
and subsequently transmit their suitably encoded sensor
readings to the FC. We model the sensor readings as time-discrete
$\mathcal{X}$-valued stochastic processes
$X_k:\Omega\times\mathcal{T}\rightarrow\mathcal{X}, (\omega,t)\mapsto
X_k[\omega,t]$, $k\in\mathcal{K}$, where
$\mathcal{X}\coloneqq[x_{\text{min}},x_{\text{max}}]$ for some given
$x_{\text{min}}<x_{\text{max}}$ is the underlying compact state space and
$\mathcal{T}$ is an at most countable set of increasingly ordered real-valued
measurement times.\footnote{Throughout the paper we skip the explicit
  designation of elementary events $\omega\in\Omega$ in the formulation of
  stochastic processes and write for example $X_k[t]$ instead of
  $X_k[\omega,t]$.} Without loss of generality let us assume that
$\mathcal{X}\subseteq\mathcal{S}\subset\mathds{R}$, where
$\mathcal{S}\coloneqq[s_{\text{min}},s_{\text{max}}]$, $s_{\text{min}}<s_{\text{max}}$, is called the \emph{sensing range}, which is
the hardware-dependent range in which the sensor elements are able to quantify
values. Finally it is assumed that the joint probability density
$p_{\ve{X}}:\mathcal{X}^K\times\mathcal{T}\rightarrow\mathds{R}_+$,
$p_{\ve{X}}(\ve{x};t)\coloneqq p_{X_1,\dots,X_K}(x_1,\dots,x_K;t)\in
C_0(\mathcal{X}^K)$, of sensor readings
$\ve{X}[t]\coloneqq(X_1[t],\dots,X_K[t])^{\sf T}$ exists, with
$C_0(\mathcal{B})$, $\mathcal{B}\subset\mathds{R}^n$, being the space of
real-valued compactly supported continuous functions over $\mathcal{B}$.
%
%
\subsection{Wireless Multiple-Access Channel}
\label{sec:w-mac}
The main contribution of this paper is a novel coding scheme that efficiently
utilizes the superposition property of the Wireless Multiple-Access Channel
(W-MAC) to compute functions of sensor readings. The W-MAC is defined as
follows.
\begin{definition}[W-MAC]
  \label{def:WMAC}
  For any transmission time $\tau\in\mathds{Z}_+$, the W-MAC is a
  map from $\mathds{C}^K$ into $\mathds{C}$ defined to be
  \begin{equation}
    \bigl(W_{1}[\tau],\dots,W_{K}[\tau]\bigr)\longmapsto\sum_{k=1}^K H_{k}[\tau]W_{k}[\tau]+N[\tau]\eqqcolon Y[\tau]\;.
    \label{eq:WMAC_def}
  \end{equation}
  Here and hereafter 
  \begin{itemize}
  \item $W_{k}[\tau]\in\mathds{C}$, $k\in\mathcal{K}$, is the transmit signal 
    of node $k$ with $\forall\tau:|W_{k}[\tau]|^2\leq
    P_{\text{max}}$, where $P_{\text{max}}>0$ is the peak power constraint on
    each node,
  \item $H_{k}[\tau]$, $k\in\mathcal{K}$, is an independent complex-valued
    flat fading process between the
    $k$\textsuperscript{th} sensor node and the FC and
  \item $N[\tau]$ is an independent complex-valued receiver
    noise process.
\end{itemize}
\end{definition}

Note that $W_{k}[\tau]$ depends on the $k$\textsuperscript{th} sensor reading
$X_k[t]\in\mathcal{X}$ at any measurement time $t\in\mathcal{T}$. If $H_{k}[\tau]\equiv 1$ and
$N[\tau]\equiv 0$, the W-MAC takes the form
\begin{equation}
  \bigl(W_{1}[\tau],\dots,W_{K}[\tau]\bigr)\longmapsto \sum_{k=1}^KW_{k}[\tau]\;,
  \label{eq:idealWMAC_def}
\end{equation}
which is referred to as the \emph{ideal W-MAC}. 
\begin{remark} \label{rem:synchronization}
  The W-MAC is a symbol-synchronous channel similar to the
  standard synchronous Code-Division Multiple-Access (CDMA) channel studied for instance in \cite{Viswanath:Anantharam:Tse:99}, \cite{Verdu:98}. We would like to emphasize, however, that the computation
  scheme proposed in this paper does not require such a synchronous channel and the only
  reason for assuming perfect synchronization is to simplify the error
  analysis in Section \ref{sec:error_analysis} and the notation throughout the
  paper.
\end{remark}
%
%
%
\subsection{Pre-processing and Post-processing Functions}
\label{sec:prepro} 
As already mentioned, the objective is not to transmit each sensor reading via
the W-MAC but rather to compute a function of these readings at the fusion
center. Throughout the paper, we use $f$ to denote the function of interest
and refer to this function as the \emph{desired function}. Obviously, given a
realization of $\ve{X}[t]$ at measurement time instance $t\in\mathcal{T}$, we have $f:\mathcal{X}^K\rightarrow\mathds{R}$, with $f\bigl(x_1[t],\dots ,x_K[t]\bigr)\eqqcolon
(f\circ\ve{x})[t]=f\bigl(\ve{x}[t]\bigr)$, where $f\bigl(\ve{x}[t]\bigr)$ is the function value which the FC attempts to
extract from the corresponding observed receive signal.

The basic idea behind the scheme for an efficient computation of desired functions proposed in this paper is to exploit the broadcast property of the W-MAC to allow the
FC to observe a superposition of signals transmitted by the
sensors. A look at Eqs. (\ref{eq:WMAC_def}) and (\ref{eq:idealWMAC_def}) shows
that the basic mathematical operation which can be naturally performed by the
W-MAC on the sensor readings is \emph{addition}. In other words, if all
sensors send their readings simultaneously over the same frequency band, then
the FC would receive a weighted sum of the sensor readings
corrupted by background noise.\footnote{In the case of an ideal W-MAC,
  the FC would observe the uncorrupted sum of sensor readings.} Now
the reader may be inclined to think that such an approach is inherently
confined for computing affine functions, which in fact is true if no additional
signal processing is carried out at the transmitters and the receiver. In this
paper, in order to overcome the restriction to affine functions, we propose to
perform some pre-processing and post-processing at the sensor nodes and the
FC, respectively. To this end, we introduce the following two
definitions.
\begin{definition}[Pre-processing Function]\label{def:pre}
  We define $\varphi_k:\mathcal{X}\rightarrow\mathds{R}$, $\varphi_k\in
  C_0(\mathcal{X})$, with $\varphi_k(x_k[t])=(\varphi_k\circ x_k)[t]$, to be a
  pre-processing function of node $k\in\mathcal{K}$.
\end{definition}
\begin{definition}[Post-processing Function]\label{def:post}
  The continuous injective function $\psi:\mathds{R}\rightarrow\mathds{R}$
  with $\psi\bigl(|y[\tau]|^2)$, where $y[\tau]$ given by (\ref{eq:WMAC_def})
  is said to be a post-processing function.\footnote{The restriction to a
    class of post-processing functions that take the squared absolute value of
    the W-MAC output as an argument is necessary because of the analog
    computation scheme proposed in Section \ref{sec:analog_over}. In general,
    the post-processing function can be defined on the set of complex
    numbers.}
\end{definition}

In order to illustrate the above definitions, it is reasonable to consider an
ideal W-MAC, in which case the objective of the pre- and post-processing functions is to transform the ideal W-MAC in
such a way that the resulting overall channel mapping from $\mathcal{X}^K$
into $\mathds{R}$ is equal to the desired function. Therefore,
$\varphi_k,k\in\mathcal{K},$ and $\psi$ are to be chosen so that
\begin{equation}
\label{eq:nomographic}
f(x_1,\dots,x_K)=\psi\Bigl(\sum\nolimits_{k\in\mathcal{K}}\varphi_k(x_k)\Bigr)\;,
\end{equation}
where $w_{k}[\tau]=\varphi_k(x_k)$ is the transmit signal of node
$k\in\mathcal{K}$ at time $\tau$.
%
%
\subsection{Functions Computable via Wireless Multiple-Access Channels}
\label{sec:computableWMAC}
\begin{figure}[t!]
	\centering
	\hspace{-14pt}
	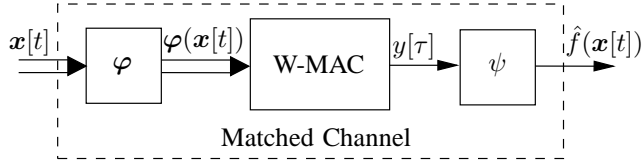
	\caption{Block diagram of the overall channel, which is matched to the
          desired function. The match results from the transformation of the
          W-MAC by the pre-processing functions
          $(\varphi_1(x_1[t]),\dots,\varphi_K(x_K[t]))^{\sf
            T}\eqqcolon\ve{\varphi}(\ve{x}[t])$ and the post-processing
          function $\psi$, respectively, which depend on the desired function
          $f$.}
	\label{fig:channel_transformation}
\end{figure}

Figure \ref{fig:channel_transformation} illustrates the functional principle
of the \emph{analog computation} scheme proposed in Section \ref{sec:analog_over}, which is referred to as the CoMAC scheme
in what follows.
Consequently, the space of all functions $\mathcal{F}(\mathcal{X}^K)\subset
C_0(\mathcal{X}^K)$ that can be computed using the analog CoMAC scheme under
the assumption of an ideal W-MAC is given by
\begin{equation}
  \mathcal{F}(\mathcal{X}^K) \coloneqq \Bigl\{f:\mathcal{X}^K\rightarrow\mathds{R}\,\Bigl|\,f\bigl(\ve{x}\bigr)=\psi\Bigl(\sum\nolimits_{k\in\mathcal{K}}\varphi_k\bigl(x_k\bigr)\Bigr)\Bigr\}\;.
  \label{eq:function_set}
\end{equation}
The space of all affine functions is clearly a subset of
$\mathcal{F}(\mathcal{X}^K)$, because any affine function can be computed if
the pre- and post-processing functions are $\varphi_k(x)=\nu_kx$, $k\in\mathcal{K}$, and $\psi(y)=a y+b$,
for some $(\nu_1,\dotsc,\nu_K)\in\mathds{R}^K$ and $a,b\in\mathds{R}$. With an
appropriate choice of the parameters, we can therefore compute any weighted
sum and, in particular, the \emph{arithmetic mean} which is of great interest
in practice. Moreover, we can easily determine the number of active nodes in a
network by letting them simultaneously transmit some constant value $c>0$ and then
post-process the received signal by means of $\psi(y)=\frac{1}{c}y$.

Now the following two questions arise immediately:
\begin{itemize}
\item[i)] Is the set of all affine
functions a \emph{proper} subset of $\mathcal{F}(\mathcal{X}^K)$? 
\item[ii)] What is exactly the function space $\mathcal{F}(\mathcal{X}^K)$ and
  how can its elements be computed?
\end{itemize}
In other words, the first question is one of whether functions other than
affine ones are members of $\mathcal{F}(\mathcal{X}^K)$ and therefore are computable using a CoMAC scheme?
The answer is obviously positive as we can easily compute the \emph{geometric
  mean} of some positive sensor readings by choosing
$\varphi_k(x)=\log_a(x),a>1,x>0$, for each $k\in\mathcal{K}$ and $\psi(y) =
a^{\frac{1}{K}y}$. Indeed, with this choice of functions, we have
$f(\ve{x})=\psi\bigl(\sum_{k\in\mathcal{K}}\varphi_k(x_k)\bigr)=(\prod_{k=1}^Kx_k)^{\frac{1}{K}}$,
where the sensor readings are positive so that $0<x_{\text{min}}\leq x_k$ for
each $k\in\mathcal{K}$. The second question in contrast is not so easy to
answer. Widely considered in wireless sensor network applications is for instance the
\emph{maximum} of sensor readings $f(\ve{x})=\max_{k\in\mathcal{K}}x_k$. It is, however, not
clear how to compute the maximum function using a CoMAC scheme. On the
positive side, the maximum function can be arbitrarily closely approximated by
a sequence of functions in $\mathcal{F}(\mathcal{X}^K)$. Indeed, it is
well-known that $\lim_{q\to\infty}\|x\|_q=f(\ve{x})=\max_{k\in\mathcal{K}}x_k$, where
$\|x\|_q=(\sum_{k=1}^Kx_k^q)^{\frac{1}{q}}\in\mathcal{F}(\mathcal{X}^K),x_k\geq
0,k\in\mathcal{K}$, and the norms can be computed when $\varphi_k(x)=x^q$, for
all $k\in\mathcal{K}$, and
$\psi(y)=y^{\frac{1}{q}}$. 

Recently, it was shown that in fact for \emph{every} multivariate function there exist pre- and post-processing functions such that they can be represented in the form (\ref{eq:nomographic}) \cite{Goldenbaum:Boche:Stanczak:11,Goldenbaum:Boche:Stanczak:12a}. The main difficulty, however, lies in a constructive characterization of $\mathcal{F}(\mathcal{X}^K)$ to determine the pre- and post-processing functions for computing arbitrary members of this space. Since an
exact constructive characterization of (\ref{eq:function_set}) is beyond the
scope of this paper, we devote our attention to the problem of computing some
functions in a robust and practically relevant manner by exploiting the
natural computational capabilities of the W-MAC.
%
%
%
%
%
%
\section{Analog Function Computation via Wireless Multiple-Access Channels}
\label{sec:analog_over}
Recent results in sensor network signal processing indicate that for many
wireless sensor network applications, an analog joint source-channel
communication architecture can be superior to widely-spread separation-based
digital approaches \cite{Gastpar:Vetterli:Dragotti:06}. This is in particular true when the processes of sensing,
computation and data transmission are highly interdependent in which case they
should be jointly considered. In order to exploit the interdependencies
by merging the processes of computation and communication, traditional analog
coding schemes require a receiver-side constructive superposition of the transmit signals from different sensor nodes in the sense of (\ref{eq:WMAC_def}) \cite{Gastpar:Vetterli:03,Bajwa:Haupt:Sayeed:Nowak:07}. However, such a perfect synchronization at the symbol and phase
level is notoriously difficult to realize in wireless networks and in
particular in large-scale wireless sensor networks \cite{Sundararaman:Buy:Kshemkalyani:05}.

Therefore, in this paper, we propose an analog computation scheme that
tolerates a \emph{coarse block-synchronization} at the FC, which is by far
easier to establish and maintain than the perfect synchronization required by
traditional approaches. The basic idea of the scheme consists in letting each sensor
node transmit a distinct sequence of complex numbers of length
$M\in\mathds{N}$ at a \emph{transmit energy} that depends on the pre-processed
sensor readings. Under some conditions and a suitable pre-processing strategy,
the received energy at the FC equals the sum of all the transmit energies
corrupted by the background noise. The coarse block-synchronization is needed to ensure a
sufficiently large overlap of different signal frames as illustrated in
Fig. \ref{fig:sequence_overlapping}. An application of an appropriately chosen
post-processing function at the receiver together with some simple arithmetic
calculations (to ensure certain estimation properties) yields then an estimate of
the desired function of the sensor readings.
%
%
%
\subsection{Computation Transmitter}
\label{sec:transmit}

\subsubsection{Data Pre-processing}
As each pre-processed sensor reading is to be encoded in transmit energy only,
it is necessary to apply a suitable bijective continuous mapping
$g_{\varphi}:[\varphi_{\text{min}},\varphi_{\text{max}}]\rightarrow[0,P_{\text{max}}]$
from the set of all pre-processed sensor readings onto the set of all
feasible transmit powers, with
$\varphi_{\text{min}}\coloneqq\min_{k\in\mathcal{K}}\inf_{x\in\mathcal{S}}\varphi_k(x)$, 
$\varphi_{\text{max}}\coloneqq\max_{k\in\mathcal{K}}\sup_{x\in\mathcal{S}}\varphi_k(x)$ and
$P_{\text{max}}$ being the transmit power constraint on each node (see Definition \ref{def:WMAC}). Note that the mapping depends on the pre-processing
functions and the sensing range and is independent of $k$, as the FC does not
have access to each individual transmit signal but only to the W-MAC output
given by (\ref{eq:WMAC_def}). We call the quantity
\begin{equation}
	P_{k}[t]\coloneqq g_{\varphi}\bigl(\varphi_k(X_k[t])\bigr) 
	\label{eq:power_def}
\end{equation}
transmit power of node $k$, and point out that it is a random variable
whenever $X_k[t]$ is random. Moreover, we have $P_{k}[t]\leq P_{\text{max}}$. Thus the information to be conveyed to the FC is
encoded in $P_{k}[t]$, for all $k\in\mathcal{K}$ and $t\in\mathcal{T}$.
%
%
%
\subsubsection{Random Sequences}
The transmit power modulates a sequence of random symbols. In what follows, we
use
\begin{equation}
  \ve{S}_k[t]\coloneqq\bigl(S_{k}[1],\dots,S_{k}[M]\bigr)^{\sf T}\in\mathds{C}^M
	\label{eq:random_sequences}
\end{equation}
to denote a sequence of transmit symbols generated by node $k$ at any measurement time
$t\in\mathcal{T}$. The symbols of the sequence are assumed to be of the form $S_{k}[m] =
\e^{i\Theta_{k}[m]}$, $m=1,\dots,M$, where
$\{\Theta_{k}[m]\}_{k,m}$ are continuous random phases that are independent
identically and uniformly distributed on $[0,2\pi)$. This implies
$\|\ve{S}_k[t]\|_2^2=M$ and a constant envelope of the transmit signal
(i.e., $|S_{k}[m]|^2=1$, for all $m,k$), which is a vital practical constraint. We have two remarks.
\begin{remark}\label{rem:discrete_phases}
  Note that the assumption of continuous random phases is not necessary for
  our CoMAC scheme to be implemented. Without loss of performance, the phases
  can take on values on any discrete subset of $[0,2\pi)$ provided that it results in a corresponding set of conjugated pairs of transmit symbols.
\end{remark}
\begin{remark}
\label{rem:NoSequenceDesign}
  Instead of optimizing the sequences assigned to different
  nodes, employing sequences with \emph{random phases} and
  constant envelope reduces the overhead for coordination and improves
  scalability when compared to systems with optimized sequences. Notice that a corresponding sequence design will probably be different from that for traditional
  asynchronous CDMA systems \cite{Verdu:98}, where the
  objective is to eliminate or reduce the mutual interference. CoMAC schemes
  in contrast have to exploit the interference for a common goal, which is the computation of functions of sensor readings.
\end{remark}
%
%
%
\subsubsection{Transmitter-side Channel Inversion}
\label{sec:channel_inversion}
If a receiver-side elimination of the impact of the fading channel may be
infeasible, we suggest that each transmitter corrects this impact by inverting
its own channel. To this end, channel state information is
necessary at each transmitter, which can be estimated from a known
pilot signal transmitted by the FC. In practical systems, the pilot
signal can also be used to wake up sensor nodes and initiate the computation
process.
With the channel state information at the nodes, each transmitter, say
transmitter $k$, inverts its channel by sending
\begin{equation}
\label{eq:WkmInput}
W_{k}[m]=\frac{\sqrt{P_k[t]}}{H_{k}[m]}S_{k}[m]
=\frac{\sqrt{P_k[t]}}{H_{k}[m]}\e^{i\Theta_{k}[m]}\;,
\end{equation}
where $W_{k}[m]$ is the W-MAC input of node $k\in\mathcal{K}$ at sequence
symbol $m$ (see also (\ref{eq:WMAC_def})). In \cite{Goldenbaum:Stanczak:10} it is shown that the
division by the channel amplitude $|H_{k}[m]|$ is sufficient so that channel
phase estimation is not necessary.

The resulting computation-transmitter structure is depicted in Fig. \ref{fig:function_value_transmitter}.

\begin{remark}\label{rem:channel_knowledge}
  Notice that any node $k$ with $P_k[t]/|H_{k}[m]|^2>P_{\text{max}}$ for some
  $m$ cannot invert its channel under the power constraint, and therefore
  must be excluded from transmissions associated with measurement time $t\in\mathcal{T}$. One possibility to mitigate
  the problem is to scale down all transmit powers by the same constant so
  that the power constraint is satisfied. Of course this impacts the
  performance in noisy channels and requires some degree of coordination. We
  are not going to dwell on this point and assume in the following that the
  set $\mathcal{K}$ is chosen such that each node can invert its own channel
  without violating the power constraint.
\end{remark}
\subsection{Computation Receiver}
\label{sec:receiver}
\begin{figure}[t!]
	\centering
	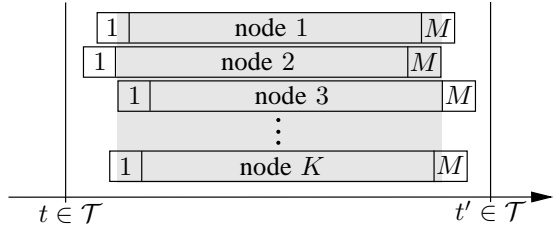
	\caption{Transmit sequences of nodes sent between measurement times $t$ and $t'$, respectively, without precise symbol- and phase-synchronization. The gray rectangle emphasizes the maximum overlapping area.}
	\label{fig:sequence_overlapping}
\end{figure}

As mentioned before (see Remark \ref{rem:synchronization}), in order to avoid
cumbersome notation and to simplify the error analysis in the next section, we
assume a perfect synchronization of signals from different nodes at the
FC. The reader however may easily verify that the proposed CoMAC scheme based
on a simple energy estimator is insensitive to the lack of synchronization
provided that a significant overlap of different signal frames is ensured as illustrated in Fig. \ref{fig:sequence_overlapping} (i.e., a coarse frame-synchronization). 
We also point out that the assumption of perfect synchronization has been widely used when analyzing asynchronous CDMA systems (see \cite{Verdu:98} and
references therein).
%
%
%
\subsubsection{Received Signal}
With this assumption in hand, the W-MAC is a memoryless channel and its output
follows with (\ref{eq:WkmInput}) from (\ref{eq:WMAC_def}) to ($1\leq
m\leq M, t\in\mathcal{T}$)
\begin{equation}
	Y[m]=\sum_{k=1}^K\sqrt{P_k[t]}S_{k}[m]+N[m]\;.
	\label{eq:Rx_signal}
\end{equation}
For any given $t\in\mathcal{T}$, we arrange the symbols in a vector
$\ve{Y}[t]\coloneqq(Y[1],\dots,Y[M])^{\sf T}\in\mathds{C}^M$ to obtain the
vector-valued W-MAC
\begin{equation}
	\ve{Y}[t] = \sum_{k=1}^K\sqrt{g_{\varphi}\bigl(\varphi_k(X_k[t])\bigr)}\,\ve{S}_k[t]+\ve{N}[t]\;,
	\label{eq:Rx_vec}
\end{equation}
where $\ve{N}[t]\coloneqq(N[1],\dots,N[M])^{\sf T}\in\mathds{C}^M$ denotes
a stationary proper complex-valued white Gaussian noise process, that is
$\ve{N}[t]\sim\mathcal{N}_{\mathds{C}}(\ve{0},\sigma_N^2\ma{I}_M)$,
$\sigma_N^2\in(0,\infty)$.
%
%
%
\subsubsection{Signal Post-processing}
The observation vector in (\ref{eq:Rx_vec}) is a basis for estimating the
desired function value $f(X_1[t],\dotsc,X_K[t])$. To this end, the receiver
first computes the received sum-energy given by
\begin{equation}
	\begin{split}
          \|\ve{Y}[t]\|_2^2{}={}&M\sum_{k=1}^K P_k[t] + \underbrace{\sum_{k=1}^K\sum_{\substack{\ell=1\\ \ell\neq k}}^K \sqrt{P_k[t]P_{\ell}[t]}\ve{S}_k[t]^{\sf H}\ve{S}_{\ell}[t]}_{\eqqcolon\Delta_1[t]\in\mathds{R}}\\
          &{+}\:\underbrace{2\sum_{k=1}^K\sqrt{P_k[t]}\,\Real\bigl\{\ve{S}_k[t]^{\sf
              H}\ve{N}[t]\bigr\}}_{\eqqcolon\Delta_2[t]\in\mathds{R}}+\underbrace{\ve{N}[t]^{\sf
              H}\ve{N}[t]}_{\eqqcolon\Delta_3[t]\in\mathds{R}_+}\;,
	\end{split}
	\label{eq:Rx_power}
\end{equation}
which can be expressed in a more compact way as
\begin{equation}
	\|\ve{Y}[t]\|_2^2=M\sum_{k=1}^K P_k[t]+\Delta[t]\;,
	\label{eq:Rx_power2}
\end{equation}
where $\Delta[t]\coloneqq\Delta_1[t]+\Delta_2[t]+\Delta_3[t]\in\mathds{R}$ is
the overall noise incorporating the three different noise sources. 

Before applying the post-processing function, the receiver must remove the
influence of the function $g_{\varphi}$, which is used to map the sensing
range on the set of feasible transmit powers. In other words, if
$\Delta[t]\equiv 0$, then an application of the post-processing function must
perfectly reconstruct the sought function value, which is expected from any
computation or transmission scheme. Now an examination of (\ref{eq:Rx_power2})
with (\ref{eq:power_def}) shows that given $g_{\varphi},\psi$ and
$\varphi_k,k\in\mathcal{K}$, we need to apply a function
$h_{\varphi}:\mathds{R}\rightarrow\mathds{R}$ to
(\ref{eq:Rx_power2}) such that
\begin{equation}
\begin{split}
  \psi\Bigl(h_{\varphi}\Bigl(M\sum_{k\in\mathcal{K}}g_{\varphi}\bigl(\varphi_k(x_k[t])\bigr)\Bigr)\Bigr)\equiv
  \psi\Bigl(\sum_{k\in\mathcal{K}}\varphi_k\bigl(x_k[t]\bigr)\Bigr)\equiv
  f\bigl(\ve{x}[t]\bigr)\in\mathcal{F}(\mathcal{X}^K)\;.
\end{split}
	\label{eq:g_h_requirement}
\end{equation}
Thus, given some pre-processing and post-processing functions, we can compute
any desired function of the form (\ref{eq:nomographic}) provided that $\Delta[t]\equiv
0$ and the pair $(g_{\varphi},h_{\varphi})$ satisfies
(\ref{eq:g_h_requirement}). The following proposition provides a necessary and
sufficient condition for the functions to fulfill (\ref{eq:g_h_requirement}).
\begin{proposition}\label{lem:affine_functions}
  Let $K\geq2$ be arbitrary. Then, (\ref{eq:g_h_requirement}) holds with
    $f$ defined by (\ref{eq:nomographic}) for some given
    $\psi,\varphi_1,\dots,\varphi_K$, if and only if $g_{\varphi}$ and
    $h_{\varphi}$ are affine functions with $h_{\varphi}\equiv
    g_{\varphi}^{-1}-c$, where the constant $c\in\mathds{R}$ depends on
    $g_{\varphi}$.
\end{proposition}
\begin{IEEEproof}
  The proof is deferred to Appendix
  \ref{app:proof_lem_affine_functions}.
\end{IEEEproof}

Examples of the data pre-processing functions and the signal post-processing
functions for the arithmetic mean and the geometric mean can be found in
Section \ref{sec:arithmetic_analysis} and Section
\ref{sec:geometric_analysis}, respectively. 
\begin{figure}[t!]
	\subfigure[]{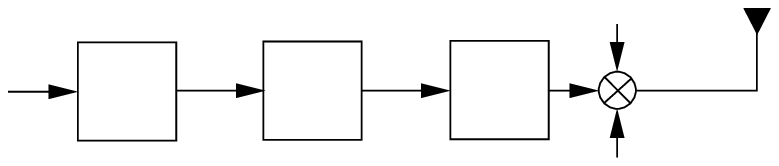\vspace{10pt}
		\label{fig:function_value_transmitter}} 
	\subfigure[]{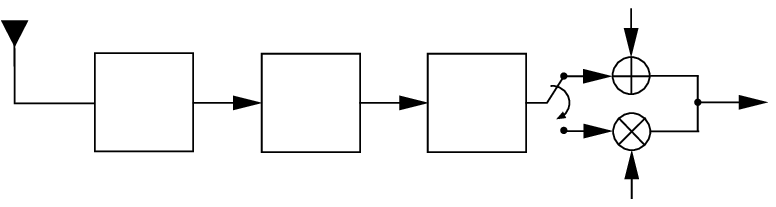
		\label{fig:function_value_receiver}}
              \caption{(a) Block diagram of the CoMAC computation-transmitter of sensor node
                $k\in\mathcal{K}$. (b) Block 
              diagram of the CoMAC computation-receiver for computing the arithemtic mean
              (switch position $1$) and the geometric mean (switch position
              $2$). Functions $h_{\varphi}$ and $\psi$ depend on the
                choice of the desired function and should be chosen according
                to the discussion in Section \ref{sec:computableWMAC} and
                Definition \ref{def:arithmetic_estimator} or Definition
                \ref{def:geometric_estimator}. For brevity, standard radio
                components (e.g., modulator, demodulator, filters) are not
                depicted.}
	\label{fig:computation_system}
\end{figure}
%
%
%
\subsection{Performance Metric}
\label{sec:perf_metric}
The performance of the CoMAC scheme is determined in terms of the
\emph{function estimation error} defined as follows.
\begin{definition}[Function Estimation Error]\label{def:recovery_error}
  Let $f\in\mathcal{F}(\mathcal{X}^K)$ be the desired function continuously
  extended onto $\mathcal{S}'$, where $\mathcal{S}'\subseteq\mathcal{S}$ is an
  appropriate subset of $\mathcal{S}$.\footnote{$\mathcal{S}'$ is introduced
    since it may be impossible to continuously extend $f$ onto the entire
    sensing range $\mathcal{S}$.}  Furthermore, let $\hat{f}$ be a
  corresponding estimate at the FC,
  $f_{\text{max}}\coloneqq\sup_{\ve{x}\in\mathcal{S}'^K}f(\ve{x})$ and
  $f_{\text{min}}\coloneqq\inf_{\ve{x}\in\mathcal{S}'^K}f(\ve{x})$. Then,
  $E\coloneqq(\hat{f}(\ve{X})-f(\ve{X}))/(f_{\text{max}}-f_{\text{min}})$ is
  said to be the \emph{function estimation error} (relative to
  $\mathcal{S}'$).
\end{definition}

Practical systems tolerate estimation errors provided that they are small
  enough. This means that $|E|\leq\epsilon$ must be satisfied for some given
  application-dependent constant $\epsilon>0$. However, in many applications,
  the requirement cannot be met permanently due to, for instance, some random
  influences. In such cases, the main figure of merit is the \emph{outage
    probability} $\Prob(|E|\geq\epsilon)$, which is the probability that the
  function estimation error is larger than or equal to $\epsilon>0$. It is
  clear that the smaller the outage probability, the higher the computation
  accuracy.

%
%
%
%
%
%
%
%
%
%
\section{Error Analysis 
}\label{sec:error_analysis}

This section is devoted to the performance analysis of the proposed CoMAC
scheme in the presence of noise. First we show that, for sufficiently large
values of $M$, the distribution of the computation noise $\Delta[t]$ can be
approximated by a normal distribution. Since the function estimation error is
strongly influenced by the post-processing function $\psi$, and with it on the
choice of the desired function $f$, we confine our
attention in subsequent subsections to two special cases of great practical importance: \emph{arithmetic
  mean} and \emph{geometric mean}. Note that these two functions are canonical
representatives of the basic arithmetic operations \emph{summation} and
\emph{multiplication}. For both cases, we define appropriate estimators by
taking into account statistical properties of the transformed overall noise
$\Delta[t]$ (transformed by $h_{\varphi}$ and $\psi$) and prove some
properties. Without loss of generality, we focus on an
arbitrary but fixed measurement time instance $t\in\mathcal{T}$ and therefore drop the
time index for brevity.
%
%
%
\subsection{Approximation of the Overall Error Distribution}\label{sec:approx_delta}%
%
%
The statistics of the overall noise in (\ref{eq:Rx_power2}) play a key role
when defining function estimators and evaluating the performance of the
proposed CoMAC scheme. Since an exact distribution of
$\Delta=\Delta_1+\Delta_2+\Delta_3$ conditioned on the sensor readings
$\ve{X}=\ve{x}$ is difficult to determine, we focus on suitable asymptotic
approximations.

To this end, let us first compute the first and second order statistical
moments of $\Delta_1$, $\Delta_2$ and $\Delta_3$. As far as $\Delta_1$ is
concerned, we have
\begin{equation}
  \Delta_1=\sum_{k=1}^K\sum_{\substack{\ell=1\\ \ell\neq k}}^K\sum_{m=1}^M\sqrt{P_kP_{\ell}}\,S_{k}^*[m]S_{\ell}[m]=2\sum_{n=1}^N\sum_{m=1}^M\sqrt{\tilde{P}_n}\,\underbrace{\cos(\Theta'_{n}[m])}_{\eqqcolon Z_{n}[m]}\;,
	\label{eq:delta1}
\end{equation}
where $N\coloneqq K(K-1)/2$, $\tilde{P}_n\coloneqq P_kP_{\ell}$ and
$\Theta'_{n}[m]\coloneqq (\Theta_{\ell}[m]-\Theta_{k}[m])\,\text{mod}\,2\pi$ the
random phase difference between nodes $k$ and $\ell$ at sequence symbol
$m$. The mapping $(k,\ell)\mapsto n$ is obtained by $n =
n(k,\ell)=\ell+(k-1)K-k(k+1)/2$, $k=1,\dots,K-1$ and $\ell=k+1,\dots,K$,
respectively.

By convolution of the densities of $\Theta_{\ell}[m]$ and $\Theta_{k}[m]$,
$\Theta'_{n}[m]$ is independent uniformly distributed over $[0,2\pi)$, for all
$n,m$. Hence, the probability density of each $Z_{n}[m]$ in
(\ref{eq:delta1}) is\footnote{Note that by the definitions, all the probability density functions and expected values in this section exists.}
\begin{equation}
	p_Z(z)=\frac{1}{\pi\sqrt{1-z^2}}\mathds{1}_{(-1,1)}(z)\;,
	\label{eq:dens_z}
\end{equation}
which is symmetric around zero. So $\forall n,m:\E\{Z_{n}[m]\}=0$ and
\begin{equation}
  \forall p_{\ve{X}}\in C_0(\mathcal{X}^K):\E\{\Delta_1\} =2\sum_{n=1}^N\sum_{m=1}^M\E\bigl\{\tilde{P}^{\frac{1}{2}}_n\bigr\}\mathbb{E}\bigl\{Z_{n}[m]\bigr\}=0\;.
  \label{eq:delta1_mean}
\end{equation}
Furthermore,
\begin{equation}
  \Var\{\Delta_1\}=4\sum_{n=1}^N\sum_{m=1}^M\E\bigl\{\tilde{P}_n\bigr\}\Var\bigl\{Z_{n}[m]\bigr\}=2M\sum_{n=1}^N\E\bigl\{\tilde{P}_n\bigr\}
	\label{eq:delta1_var}
\end{equation}
since $\forall m,n\neq n':\Cov\{Z_{n}[m],Z_{n'}[m]\}=0$ and $\forall
m,n:\Var\{Z_{n}[m]\}=1/2$, where the latter can be concluded by considering
(\ref{eq:dens_z}). As for the second error term $\Delta_2$, we have
\begin{equation}
  \Delta_2=2\sum_{k=1}^K\sqrt{P_k}\,\Real\bigl\{\ve{S}_k^{\sf H}\ve{N}\bigr\}=2\sum_{k=1}^K\sum_{\ell=1}^{2M}\sqrt{P_k}\,U_{k\ell}N'_{\ell}
  \label{eq:delta2}
\end{equation}
where for any odd $\ell$, $U_{k\ell}\coloneqq\cos(\Theta_{k}[m])$,
$N'_{\ell}\coloneqq\Real\{N[m]\}$ and $U_{k\ell}\coloneqq\sin(\Theta_{k}[m])$,
$N'_{\ell}\coloneqq\Imag\{N[m]\}$, for any even $\ell$ ($m=1,\dots,M$). Notice
that $\forall\ell:N'_{\ell}\sim\mathcal{N}_{\mathds{R}}(0,\frac{1}{2}\sigma_N^2)$
and the probability density function of $U_{k\ell}$ is given by
(\ref{eq:dens_z}). Because $N'_{\ell}$ and $U_{k\ell}$ are zero mean and
independent for all $k,\ell$, it follows for the expectation value
\begin{equation}
  \forall p_{\ve{X}}\in C_0(\mathcal{X}^K):\E\{\Delta_2\} = 2\sum_{k=1}^K\sum_{\ell=1}^{2M}\E\bigl\{\sqrt{P_k}\bigr\}\,\E\{U_{k\ell}\}\E\{N'_{\ell}\}=0\;.
	\label{eq:delta2_mean}
\end{equation}
Arguing along similar lines as in the case of $\Delta_1$, the variance of
$\Delta_2$ can be easily shown to be
\begin{equation}
  \Var\{\Delta_2\} =
  4\sum_{k=1}^K\sum_{\ell=1}^{2M}\E\{P_k\}\Var\{U_{k\ell}\}\Var\{N'_{\ell}\}
=2M\sigma_N^2\sum_{k=1}^K\E\{P_k\}\;.
	\label{eq:delta2_var}
\end{equation}
%
%

Since $\Delta_3=\sum_m|N[m]|^2\sim\chi_{2M}^2$, we finally conclude
$\E\{\Delta_3\}=M\sigma_N^2$ and
\begin{equation}
	\Var\{\Delta_3\}=M\sigma_N^4\;.
	\label{eq:delta3_var}
\end{equation}
\begin{lemma}\label{lem:uncorrelated}
  $\Delta_1$, $\Delta_2$ and $\Delta_3$ are mutually orthogonal (in the
  Hilbert space of random variables with the inner product defined to be
  $\langle\Delta_j,\Delta_{j'}\rangle\equiv\E\{\Delta_j\Delta_{j'}\}$) for all
  $p_{\ve{X}}\in C_0(\mathcal{X}^K)$.
\end{lemma}
\begin{IEEEproof}
  Since the sensor readings, the sequence symbols and the noise are mutually
  independent random variables with $\forall m:\E\{N[m]\}=0$, a straightforward
  calculation of the covariances between $\Delta_1$ and $\Delta_2$ as well as
  between $\Delta_2$ and $\Delta_3$ proves the lemma.
\end{IEEEproof}

The above derivations show that $\forall p_{\ve{X}}\in
C_0(\mathcal{X}^K):\E\{\Delta\}=M\sigma_N^2$, while, by Lemma
\ref{lem:uncorrelated}, the variance of $\Delta$ is the sum of the variances
(\ref{eq:delta1_var}), (\ref{eq:delta2_var}) and (\ref{eq:delta3_var}). Thus,
\begin{equation}
  \sigma_{\Delta}^2\coloneqq\Var\{\Delta\}
  =2M\sum_{n=1}^N\E\bigl\{\tilde{P}_n\bigr\}
  +2M\sigma_N^2\sum_{k=1}^K\E\bigl\{P_k\bigr\}+M\sigma_N^4
	\label{eq:Delta_var}
\end{equation}
and note that when conditioned on $\ve{X}=\ve{x}$, the variance in
(\ref{eq:Delta_var}) yields
\begin{equation}
  \sigma_{\Delta|\ve{x}}^2\coloneqq\E\{(\Delta-M\sigma_N^2)^2|\ve{X}=\ve{x}\}
  =2M\sum_{n=1}^N\tilde{p}_n+2M\sigma_N^2\sum_{k=1}^Kp_k+M\sigma_N^4\;.
	\label{eq:Delta_var_cond}
\end{equation}

As mentioned in the introduction to this section, we were not able to find out
the exact distribution of the overall noise $\Delta$, which includes various
terms with different distributions. However, since the number of summands
$J\coloneqq K(K-1)M/2+2KM+2M$ in the definition of $\Delta$ is already
relatively large for small values of $K$ and $M$, we argue that it is
well-founded to invoke the central limit theorem so as to approximate the
conditional distribution by a normal distribution. The following proposition proves the corresponding convergence as $M\to\infty$.
\begin{proposition}\label{prop:central_limit_theorem}
  Let $\Delta|\ve{x}$ be the overall noise according to (\ref{eq:Rx_power})
  and (\ref{eq:Rx_power2}) conditioned on the sensor readings $\ve{X}=\ve{x}$
  with $\E\{\Delta\,|\,\ve{X}=\ve{x}\}=M\sigma_N^2$, $0<\sigma_N^2<\infty$, and
  $\sigma_{\Delta|\ve{x}}^2$ as defined in (\ref{eq:Delta_var_cond}). Then,
  for any fixed $K,P_{\text{max}}<\infty$ and a compact set $\mathcal{X}$, we
  have
  \begin{equation}
    \forall\ve{x}\in\mathcal{X}^K:\frac{\Delta|\ve{x}-M\sigma_N^2}{\sigma_{\Delta|\ve{x}}}\stackrel{\text{d}}{\longrightarrow}\mathcal{N}_{\mathds{R}}(0,1)
    \label{eq:central_limit}
  \end{equation}
  as $M\rightarrow\infty$, where $\stackrel{\text{d}}{\longrightarrow}$
  denotes the convergence in distribution.
\end{proposition}
\begin{IEEEproof} 
  Since the sum terms of $\Delta|\ve{x}$ are neither identically distributed
  nor independent, the convergence to a normal distribution is not clear. Let
  us therefore rearrange the sum to obtain:
  %
  \begin{equation*}
    \begin{split}
    \Delta|\ve{x}{}={}&\Delta_1|\ve{x}+\Delta_2|\ve{x}+\Delta_3
    =\sum_{n=1}^N\sum_{m=1}^M\sqrt{\tilde{p}_n}\cos (\Theta'_{n}[m])+2\sum_{k=1}^K\sum_{\ell=1}^{2M}\sqrt{p_k}U_{k\ell}N'_{\ell}+\sum_{m=1}^M|N[m]|^2\\
    {}={}&\sum_{m=1}^M\Biggl[\sum_{n=1}^N\sqrt{\tilde{p}_n}\cos (\Theta'_{n}[m])
+\sum_{k=1}^K\Bigl(\Real\{N[m]\}\cos(\Theta_{k}[m])\cdots\\
&\hspace{128.5pt}{}+{}\Imag\{N[m]\}\sin(\Theta_{k}[m])\Bigr)+|N[m]|^2\Biggr]
    =\sum_{m=1}^M\Lambda_m\;.
    \end{split}
  \end{equation*}
  %
  This makes clear that $\Lambda_m$, $m=1,\dots,M$, are independent and
  identically distributed nondegenerate (i.e., $\Var\{\Lambda_1\}>0$) random
  variables. Moreover, for any $K,P_{\text{max}},\sigma_N^2<\infty$ and a
  compact set $\mathcal{X}$,
  $\E\{\Lambda_1^2\,|\,\ve{X}=\ve{x}\}=2\left(\sum_{n=1}^N\tilde{p}_n+\sigma_N^2\sum_{k=1}^Kp_k+\sigma_N^4\right)$
  is finite. Hence the proposition follows from Theorem 3 in
  \cite[p.\,326]{Shiryaev:96} with (\ref{eq:Delta_var_cond}) and
  $\E\{\Delta\,|\,\ve{X}=\ve{x}\}=M\sigma_N^2$.
\end{IEEEproof}

Since Proposition \ref{prop:central_limit_theorem} implies the uniform
convergence of the sequence of distribution functions associated with
$\{\Delta|\ve{x}\}_{M\in\mathds{N}}$, we can conclude that the distribution of
$\Delta|\ve{x}$ can be approximated by a normal distribution provided that $M$
is sufficiently large. This is summarized in a corollary.
\begin{corollary}\label{cor:degenerated}
  If $M$ is sufficiently large, $\Delta|\ve{x}$ is close to
  $\tilde{\Delta}|\ve{x}\sim\mathcal{N}_{\mathds{R}}(M\sigma_N^2,\sigma_{\Delta|\ve{x}}^2)$
  in distribution.
\end{corollary}
We point out that determining the convergence rate is beyond the scope of this paper, extensive
numerical experiments (see Section \ref{sec:num_examples}) suggest that the approximation stated in Corollary
\ref{cor:degenerated} is justified already for small values of $M$ and most
cases of practical interest.
%
%
%
%
\subsection{Arithmetic Mean Analysis}\label{sec:arithmetic_analysis}
%
%
First, we define a suitable \emph{arithmetic mean} estimator based on the
observation of the channel output energy $\|\ve{Y}\|_2^2$ given by
(\ref{eq:Rx_power2}). Subsequently, we analyze the outage probability under
the proposed estimator.

\begin{definition}[Arithmetic Mean Estimate]
  \label{def:arithmetic_estimator}
  Let $f$ be the desired function ``arithmetic mean'' and let the expected value $\E\{\psi(\Delta_3/(M\alpha_{\text{arit}}))\}$ be known to the FC. Then, given $M$, the estimate
  $\hat{f}_M(\ve{X})$ of $f(\ve{X})$ is defined to be
  \begin{equation}
    \hat{f}_M(\ve{X})\coloneqq\psi\bigl(h_{\varphi}(\|\ve{Y}\|_2^2)\bigr) 
    -\E\left\{\psi\bigl(\Delta_3/(\alpha_{\text{arit}}M)\bigr)\right\}\;.
    \label{eq:arith_estimator}
  \end{equation}
  Assuming $M\sum_kg_{\varphi}(\varphi_k(x_k))=M\sum_kp_k\eqqcolon z$ and
  $\alpha_{\text{arit}}\coloneqq
  \frac{P_{\text{max}}}{s_{\text{max}}-s_{\text{min}}}$, we have
  \begin{itemize}
  \item \emph{Data pre-processing}: $\forall
    k:\varphi_k(x)=x,g_{\varphi}(x)=\alpha_{\text{arit}}(x-s_{\text{min}})$,
    $\varphi_{\text{min}}=s_{\text{min}}$,
    $\varphi_{\text{max}}=s_{\text{max}}$,
  \item \emph{Signal post-processing}: $\psi(x)=x/K,x\in\mathds{R}$,
    $h_{\varphi}(z)=\frac{1}{M\alpha_{\text{arit}}}z+Ks_{\text{min}}$.
\end{itemize}
\end{definition}

Now, we prove two propositions to show that the arithmetic mean estimator of
Definition \ref{def:arithmetic_estimator} provides two most desired
properties: unbiasedness and consistency. The resulting computation-receiver is depicted in Fig. \ref{fig:function_value_receiver} with the
switch in position 1.
\begin{proposition}\label{prop:arith_bias}
  The function value estimator of Definition \ref{def:arithmetic_estimator} is
  \emph{unbiased}, that is, we have
  $\forall\ve{x}\in\mathcal{X}^K:\mathbb{E}\{\hat{f}_M(\ve{X})\,|\,\ve{X}=\ve{x}\}=f(\ve{x})$.
\end{proposition}
\begin{IEEEproof}
  With the definitions introduced in Section \ref{sec:computableWMAC} and
  Definition \ref{def:arithmetic_estimator} in mind, we can write
  (\ref{eq:arith_estimator}) as $\hat{f}_M(\ve{X}) =
  f(\ve{X})+\frac{1}{\alpha_{\text{arit}}KM}(\Delta-M\sigma_N^2)$. From this,
  it follows that
  $\E\{\hat{f}_M(\ve{X})\,|\,\ve{X}=\ve{x}\}=f(\ve{x})+\frac{1}{\alpha_{\text{arit}}KM}(\E\{\Delta\,|\,\ve{X}=\ve{x}\}-M\sigma_N^2)=f(\ve{x})$. So
  the proposition follows since
  $\forall\ve{x}\in\mathcal{X}^K:\E\{\Delta\,|\,\ve{X}=\ve{x}\}=\E\{\Delta_3\}=M\sigma_N^2$.
\end{IEEEproof}
\begin{proposition}\label{prop:arith_consistent}
  Let $K,P_{\text{max}},\sigma_N^2<\infty$ be arbitrary but fixed, and let
  $\{\hat{f}_M\}_{M\in\mathds{N}}$ be a sequence of estimators
  (\ref{eq:arith_estimator}). Then, the arithmetic mean estimator
  $\hat{f}$ of Definition \ref{def:arithmetic_estimator} is \emph{consistent},
  that is
  $\forall\epsilon>0:\lim_{M\rightarrow\infty}\Prob(|\hat{f}_M-f|\geq\epsilon)=0$.
\end{proposition}
\begin{IEEEproof}
  Let $c\coloneqq1/(\alpha_{\text{arit}}K)>0$ and $\epsilon>0$ be arbitrary and fixed. By
  the preceding proof, we know that
  $E_M\coloneqq\hat{f}_M-f=\frac{c}{M}(\Delta-\E\{\Delta_3\})$. Hence, as
  $\E\{\Delta\}=\E\{\Delta_3\}$, we obtain
  \setlength{\arraycolsep}{0.0em}%
  \begin{eqnarray}
	\Prob(|E_M|\geq\epsilon)&{}={}&\Prob(E_M^2\geq\epsilon^2)
      =\Prob\bigl(c^2 M^{-2}(\Delta-\E\{\Delta_3\})^2\geq\epsilon^2\bigr)\nonumber\\
      &{}\leq{}&c^2(M\epsilon)^{-2}\E\bigl\{(\Delta-\E\{\Delta\})^2\bigr\}
      =c^2(M\epsilon)^{-2}\Var\{\Delta\}\;,
    \label{eq:Ubound_Markov}
   \end{eqnarray}
  \setlength{\arraycolsep}{5pt}%
  where we used Markov's inequality \cite[p.\,47]{Shiryaev:96} (also
  called Chebyshev's inequality). By (\ref{eq:Delta_var}), we have for $K,P_{\text{max}},\sigma_N^2<\infty$ that $\Var\{\Delta\}\in\mathcal{O}(M)$ so that the right-hand side of the above
  inequality goes to zero as $M$ tends to infinity. Since $\epsilon>0$ is
  arbitrary, this completes the proof.
\end{IEEEproof}

Since the upper bound in (\ref{eq:Ubound_Markov}) typically provides
  rather loose bounds for finite values of $M$, we cannot use it to
  approximate the outage probability. It turns out that a better approach is
  to invoke Proposition \ref{prop:central_limit_theorem} and approximate $\Prob(|E|\geq\epsilon)$ by using a transformed normal distribution. Note that as
$f_{\text{max}}=\sup_{\ve{x}\in\mathcal{S}^K}f(\ve{x})=s_{\text{max}}$ and
$f_{\text{min}}=\sup_{\ve{x}\in\mathcal{S}^K}f(\ve{x})=s_{\text{min}}$ with
$f$ being continuously extended onto $\mathcal{S}$, we have
\begin{equation}
  E|\ve{x}=\bigl(\hat{f}_M(\ve{x})-f(\ve{x})\bigr)\big/(s_{\text{max}}-s_{\text{min}})
  =(\Delta|\ve{x}-M\sigma_N^2)/\alpha'_{\text{arit}}\;,
  \label{eq:error_arith}
\end{equation}
where $\alpha'_{\text{arit}}\coloneqq MKP_{\text{max}}$. 

The Mann-Wald theorem \cite[p.\,356]{Shiryaev:96}\label{th:mann_wald} guarantees that, for any real
continuous mapping $h=h(x)$, one has
$h(X_n)\stackrel{\text{d}}{\rightarrow}h(X)$ whenever
$X_n\stackrel{\text{d}}{\rightarrow}X$. We can therefore conclude from Corollary \ref{cor:degenerated} that for
sufficiently large values of $M$, $E|\ve{x}$ in (\ref{eq:error_arith}) can be
approximated by a random variable $\tilde{E}|\ve{x}\sim\mathcal{N}_{\mathds{R}}(0,\tfrac{\sigma_{\Delta|\ve{x}}^2}{{\alpha'}_{\text{arit}}^2})$
with conditional distribution function
$P_{\tilde{E}}(e|\ve{x})\coloneqq\Prob(\tilde{E}\leq e|\ve{X}=\ve{x})=\tfrac{1}{2}[1+\erf(\tfrac{\alpha'_{\text{arit}}e}{{\sigma_{\Delta|\ve{x}}\sqrt{2}}})]$,
$e\in\mathds{R}$.  

Since the absolute value is also
continuous and
$\Prob(|\tilde{E}|\geq\epsilon|\ve{X}=\ve{x})=1-P_{\tilde{E}}(\epsilon|\ve{x})+P_{\tilde{E}}(-\epsilon|\ve{x})$
for any $\epsilon>0$, we obtain for sufficiently large $M$,
\begin{IEEEeqnarray}{rCl}
  \Prob(|E|\geq \epsilon)\approx\Prob(|\tilde{E}|\geq\epsilon)&=&\int_{\mathcal{X}^K}\!\Prob(|\tilde{E}|\geq \epsilon\,|\,\ve{X}=\ve{x})p_{\ve{X}}(\ve{x})\,d\ve{x}\IEEEnonumber\\
  &=&\int_{\mathcal{X}^K}\!\erfc\bigl(\alpha'_{\text{arit}}\epsilon/(2\sigma_{\Delta|\ve{x}}^2)^{\frac{1}{2}}\bigr)p_{\ve{X}}(\ve{x})\,d\ve{x}\;,
	\label{eq:prob_approx_gauss}
\end{IEEEeqnarray}
where we used the fact that $\erf(-x)=-\erf(x)$ for all $x\in\mathds{R}$. 
%
%
%
%
\subsection{Geometric Mean Analysis}\label{sec:geometric_analysis}
%
As in the preceding subsection, we first define an estimator for the desired function
\emph{geometric mean} including the required data pre-processing and signal
post-processing functions.
\begin{definition}[Geometric Mean Estimate]\label{def:geometric_estimator}
  Let $f$ be the desired function ``geometric mean'' as defined in Section
  \ref{sec:computableWMAC}, and let the expected value
  $\E\{\psi(\Delta_3/\alpha_{\text{geo}})\}$ be known to the FC (see Lemma
  \ref{lem:beta} below). Then, given $M$, the estimate $\hat{f}_M(\ve{X})$ of $f(\ve{X})$ is defined to be
  \begin{equation}
    \hat{f}_M(\ve{X})\coloneqq\frac{\psi\bigl(h_{\varphi}(\|\ve{Y}\|_2^2)\bigr)}{\E\left\{\psi\bigl(\Delta_3/(\alpha_{\text{geo}}M)\bigr)\right\}}
    =f\bigl(\ve{X}\bigr)\frac{\psi\bigl(\Delta/(\alpha_{\text{geo}}M)\bigr)}{\E\left\{\psi\bigl(\Delta_3/(\alpha_{\text{geo}}M)\bigr)\right\}}\,.
    \label{eq:f_estimate}
  \end{equation}
  Assuming $M\sum_kg_{\varphi}(\varphi_k(x_k))=M\sum_kp_k\eqqcolon z$ and
  $\alpha_{\text{geo}}\coloneqq
  \frac{P_{\text{max}}}{\log_a(s_{\text{max}})-\log_a(s')}$, we have
  \begin{itemize}
  \item \emph{Data pre-processing}: If $s_{\text{min}}\leq0$, choose an
    arbitrary but fixed $s'$ such that $0<s'\leq x_{\text{min}}<
    s_{\text{max}}$ and otherwise $s'=s_{\text{min}}$. Then, $\forall
    k:\varphi_k(x)=\log_a(x)$, $a>1$, $\varphi_{\text{min}}=\log_a(s')$
    and $\varphi_{\text{max}}=\log_a(s_{\text{max}})$, and $\forall
    k:g_{\varphi}(\log_a(x))=\alpha_{\text{geo}}(\log_a(x)-\log_a(s'))$.
  \item \emph{Signal post-processing}: $\psi(x)=a^{x/K},x\in\mathds{R}$,
    $h_{\varphi}(z)=\frac{1}{M\alpha_{\text{geo}}}z+K\log_a(s')$.
  \end{itemize}
\end{definition}

The resulting computation-receiver is shown in
Fig. \ref{fig:function_value_receiver} with the switch in position 2. As
mentioned, our estimator requires the knowledge of
$\E\{\psi(\Delta_3/(\alpha_{\text{geo}}M))\}$, which is explicitly given in
part (i) of the following lemma. Part (ii) is used in the proof of Proposition
\ref{prop:properties}.
\begin{lemma}\label{lem:beta}
  Let $a>1$ be given and fixed, and let $\alpha_{\text{geo}}$ be as in
  Definition \ref{def:geometric_estimator}. Suppose that
  $\sigma_N^2\log_{\e}(a)<\alpha_{\text{geo}} KM$. Then
  \begin{itemize}
  \item[(i)]
    $\lambda_M\coloneqq\E\left\{\psi\bigl(\Delta_3/(\alpha_{\text{geo}}M)\bigr)\right\}=\left(\frac{\alpha_{\text{geo}}
        KM}{\alpha_{\text{geo}} KM-\sigma_N^2\log_{\e}(a)}\right)^M$
  \item[(ii)] $\displaystyle\lim_{M\rightarrow\infty}\lambda_M
    =\e^{\frac{\sigma_N^2\log_{\e}(a)}{\alpha_{\text{geo}}K}}$\;.
  \end{itemize}
\end{lemma}
\begin{IEEEproof}
  The proof is deferred to Appendix \ref{app:proof_lem_beta}.
\end{IEEEproof}

We point out that the expected value $\lambda_M$ exists if
$\sigma_N^2\log_{\e}(a)<\alpha_{\text{geo}} KM$ holds, which is usually
fulfilled in practical situations and therefore assumed in what follows.
%
With the estimator of Definition \ref{def:geometric_estimator}, the fuction estimation
error conditioned on the sensor readings $\ve{X}=\ve{x}$ becomes
\begin{equation}
	E|\ve{x}= 
        \frac{1}{\gamma(\ve{x})}\Xi|\ve{x}-\beta(\ve{x})
        =\beta(\ve{x})\left(\frac{\Xi|\ve{x}}{\lambda_M}-1\right)\;,
	\label{eq:error_geo1}
\end{equation}	
where we used the following notation: $f_{\text{max}}=s_{\text{max}}$,
$f_{\text{min}}=s'$ ($0<s'\leq x_{\text{min}}$), $\beta(\ve{x})\coloneqq f(\ve{x})/(s_{\text{max}}-s')$,
$\gamma(\ve{x})\coloneqq\lambda_M/\beta(\ve{x})$
and $\Xi|\ve{x}\coloneqq\psi(\Delta|\ve{x}/(\alpha_{\text{geo}}M))$.

Note that the estimator of Definition \ref{def:geometric_estimator} is not
necessarily unbiased but it offers the advantage of a simple implementation in
practical systems. In contrast, the estimator
\begin{equation}
  \hat{f}_M(\ve{X})
  =\psi\bigl(h_{\varphi}(\|\ve{Y}\|_2^2)\bigr)\E\left\{\psi\bigl(\Delta/(\alpha_{\text{geo}}M)\bigr)\right\}^{-1}
  \label{eq:f_estimate_un}
\end{equation}
is unbiased but not applicable in practice, because in contrast to the expected value in (\ref{eq:f_estimate}) depends $\E\left\{\psi\bigl(\Delta/(\alpha_{\text{geo}}M)\bigr)\right\}$ in
(\ref{eq:f_estimate_un}) on the overall noise, and thus on the
distribution of the sensor readings, which is usually unknown at the FC.

Although the estimator is not unbiased, the following proposition shows that
it is (weakly) consistent, and therefore asymptotically unbiased. 
%
\begin{proposition}\label{prop:properties}
  For any fixed $K,P_{\text{max}},\sigma_N^2<\infty$, the geometric
  mean estimator proposed in Definition \ref{def:geometric_estimator} is
  \emph{consistent}.
\end{proposition}
\begin{IEEEproof}
  Let $K,P_{\text{max}},\sigma_N^2<\infty$ and $\epsilon>0$ be arbitrary and
  fixed. Let $\{\hat{f}_M\}_{M\in\mathds{N}}$ be the sequence of estimators given by (\ref{eq:f_estimate}). We show that the outage probability 
  $\Prob(|E|\geq\epsilon)\to0$ as $M\to\infty$. To this end, consider
  $\Prob(|E|\ve{x}|\geq\epsilon)\coloneqq\Prob(|E|\geq\epsilon|\ve{X}=\ve{x})$
  for any $\ve{x}\in\mathcal{X}^K$, and note that
  $f(\ve{x})>0,\beta(\ve{x})>0,\lambda_M>0$ and $\Xi_{\ve{x}}\coloneqq\Xi|\ve{x}>0$. 
  By (\ref{eq:error_geo1}), we have
  $\Prob(|E|\ve{x}|\geq\epsilon)=\Prob(\Xi_{\ve{x}}/\lambda_M\geq1+\epsilon/\beta(\ve{x}))
  +\Prob(1-\Xi_{\ve{x}}/\lambda_M\geq\epsilon/\beta(\ve{x}))$. An application
  of Markov's inequality \cite[p.\,47]{Shiryaev:96} yields an upper bound
  on the first sum term:
  \begin{equation}
    \label{eq:geo_uppebound_markov}
      \Prob(\Xi_{\ve{x}}/\lambda_M\geq1+\epsilon/\beta(\ve{x}))=
      \Prob\Bigl(\log_{\e}\Bigl(\tfrac{\Xi_{\ve{x}}}{\lambda_M}\Bigr)\geq\log_{\e}\bigl(1+\epsilon/\beta(\ve{x})\bigr)\Bigr)
      \leq\frac{\E\{\log_{\e}(\Xi_{\ve{x}})\}-\log_{\e}(\lambda_M)}{\log_{\e}\bigl(1+\epsilon/\beta(\ve{x})\bigr)}\;.
  \end{equation}
  By (ii) of Lemma \ref{lem:beta}, we have
  $\lim_{M\to\infty}\log_{\e}(\lambda_M)=\log_{\e}(\lim_{M\to\infty}\lambda_M)=\tfrac{\sigma_N^2\log_{\e}(a)}{\alpha_{\text{geo}}K}$.
  By the results on the distribution functions of random variables that are
  functions of other random variables \cite[pp.\,239--240]{Shiryaev:96}, we
  obtain
  $\E\{\log_{\e}(\Xi_{\ve{x}})\}=\tfrac{\log_{\e}(a)}{K}\E\bigl\{\tfrac{\Delta|\ve{x}}{\alpha_{\text{geo}}M}\bigr\}
  =\tfrac{\sigma_N^2\log_{\e}(a)}{\alpha_{\text{geo}}K}$, where we used
  $\E\{\Delta\,|\,\ve{X}=\ve{x}\}=M\sigma_N^2$ in the last step. Combining the
  results shows that the upper bound in (\ref{eq:geo_uppebound_markov}) tends
  to zero as $M\to\infty$. As for
  $\Prob(1-\Xi_{\ve{x}}/\lambda_M\geq\epsilon/\beta(\ve{x}))$, note that we
  can focus on $\epsilon/\beta(\ve{x})<1$ since
  $\Xi_{\ve{x}}/\lambda_M>0$. With this in hand, we have
  $\Prob(1-\Xi_{\ve{x}}/\lambda_M\geq\epsilon/\beta(\ve{x}))=\Prob(\lambda_M/\Xi_{\ve{x}}\geq1/(1-\epsilon/\beta(\ve{x})))$. Proceeding
  essentially along the same lines as above shows that this probability goes
  to zero with $M\to\infty$. Now, by compactness of $\mathcal{X}$ and Theorem
  3 or Theorem 4 of \cite[p.\,188]{Shiryaev:96}, we have
  $\lim_{M\to\infty}\Prob(|E|\geq\epsilon)=\lim_{M\to\infty}\E\{\Prob(|E_M|\geq\epsilon\,|\,\ve{X}=\ve{x})\}
  =\E\{\lim_{M\to\infty}\Prob(|E|\geq\epsilon\,|\,\ve{X}=\ve{x})\}\to0$.
\end{IEEEproof}
\begin{remark}\label{rem:accuracy}
  Notice that the proposition implies that the proposed geometric mean estimator (\ref{eq:f_estimate}) is asymptotically unbiased, that is, we have
  $\lim_{M\rightarrow\infty}\E\{\hat{f}(\ve{X})\,|\,\ve{X}=\ve{x}\}=f(\ve{x})$. As a consequence, the proposed estimator (\ref{eq:f_estimate}) is asymptotically equivalent to (\ref{eq:f_estimate_un}).
\end{remark}

Unfortunately, $\Prob(|E|\geq\epsilon)$ cannot be exactly evaluated because we
are not able to determine the distribution function of
$|E|=|\gamma(\ve{X})^{-1}\Xi-\beta(\ve{X})|$. For this reason, as in the
preceding subsection, we approximate the distribution of $\Xi|\ve{x}$ by a
transformed normal distribution since in contrast to the arithmetic mean case depends
$\Xi|\ve{x}$ nonlinearly on the conditioned overall noise
$\Delta|\ve{x}$.

\begin{lemma}\label{lem:Xi_phi}
  Let $K<\infty$, $\mathcal{X}$ be compact and $M$ sufficiently large. Then,
  $\Xi|\ve{x}$ can be approximated by a random variable
  $\tilde{\Xi}|\ve{x}\sim\mathcal{LN}(\mu_{\Xi},\sigma_{\Xi|\ve{x}}^2)$, where
  $\mu_{\Xi}=\sigma_N^2\log_{\e}(a)/(\alpha_{\text{geo}} K)$ and
  $\sigma_{\Xi|\ve{x}}^2=\sigma_{\Delta|\ve{x}}^2(\log_{\e}(a))^2/(\alpha_{\text{geo}}K)^2$,
  respectively.
\end{lemma}
\begin{IEEEproof}
  The proof can be found in Appendix \ref{app:proof_lem_log1}.
\end{IEEEproof}

With Lemma \ref{lem:Xi_phi} in hand, we are now in a position to prove the
main result of this section.
\begin{proposition}\label{prop:prob_abs_cond}
  Consider the proposed geometric mean estimator (\ref{eq:f_estimate}) and
  suppose that $E$ is the corresponding function estimation
  error. Let $\mu_{\Xi}$ and $\sigma^2_{\Xi|\ve{x}}$ be given by Lemma
  \ref{lem:Xi_phi}, and let $\beta(\ve{x}),\gamma(\ve{x})>0$ be as defined in
  (\ref{eq:error_geo1}). Then, for $M$ sufficiently large, the outage
  probability $\Prob(|E|\geq\epsilon)$, $\epsilon> 0$, can be approximated by
  \begin{equation}
    \Prob(|E|\geq\epsilon)\approx\Prob(|\tilde{E}|\geq\epsilon)=\int_{\mathcal{X}^K}\!\Prob(|\tilde{E}|\geq \epsilon|\ve{X}=\ve{x})p_{\ve{X}}(\ve{x})\,d\ve{x}
    \label{eq:prob_end}
  \end{equation}
  with
  \begin{equation}
    \Prob(|\tilde{E}|\geq \epsilon|\ve{X}=\ve{x})=
    \begin{cases}
      \frac{1}{2}\Big[2+\erf\Big(\frac{\log_{\e}(\rho^-(\ve{x},\epsilon))-\mu_{\Xi}}{\sqrt{2}\,\sigma_{\Xi|\ve{x}}}\Big)-\erf\Big(\frac{\log_{\e}(\rho^+(\ve{x},\epsilon))-\mu_{\Xi}}{\sqrt{2}\,\sigma_{\Xi|\ve{x}}}\Big)\Big],&0<\epsilon <\beta(\ve{x})\\
      \frac{1}{2}\erfc\Big(\frac{\log_{\e}(\rho^+(\ve{x},\epsilon))-\mu_{\Xi}}{\sqrt{2}\,\sigma_{\Xi|\ve{x}}}\Big),&\beta(\ve{x})\leq\epsilon <\infty	
    \end{cases},
    \label{eq:prob_sub}
  \end{equation}
  where $\rho^-(\ve{x},\epsilon)\coloneqq\gamma(\ve{x})(\beta(\ve{x})-\epsilon)$ and $\rho^+(\ve{x},\epsilon)\coloneqq\gamma(\ve{x})(\beta(\ve{x})+\epsilon)$, respectively.
\end{proposition}
\begin{IEEEproof}
  The proof is deferred to Appendix \ref{app:proof_prop_prob}.
\end{IEEEproof}

In Section \ref{sec:accuracy}, we choose a particular density
$p_{\ve{X}}(\ve{x})$ and evaluate (\ref{eq:prob_end}) numerically to indicate
the accuracy of the approximation for different network parameters.
%
%
%
%
%
%
\section{Numerical Examples} \label{sec:num_examples}
The objective of this section is twofold. First, we show in Section
\ref{sec:accuracy} that the approximations of Section \ref{sec:error_analysis}
are very accurate, and second, we compare in Section \ref{sec:comparison_tdma}
the proposed analog CoMAC scheme with a TDMA-based scheme to indicate the
huge potential for performance gains in typical sensor network operating
points.

As a basis, we consider a classical environmental monitoring scenario in which
the FC is interested in the arithmetic mean or geometric mean of temperature
measurements taken by a number of sensor nodes distributed over some
geographical area. We assume that all nodes are equipped with a low-power
temperature sensor supporting a typical sensing range
$\mathcal{S}=[-55\,^\circ\text{C},130\,^\circ\text{C}]$
\cite{STMicroelectronics:09}.
%
%
%
\subsection{Approximation Accuracy}\label{sec:accuracy}
%
%
%
To assess the accuracy of the approximated distributions, we consider two
scenarios: one where the FC estimates the arithmetic mean, and one where the geometric mean is desired. We compare (\ref{eq:prob_approx_gauss})
and (\ref{eq:prob_end}) with Monte Carlo evaluations of the outage probability
$\Prob(|E|\geq\epsilon)$ based on $10\cdot10^3$ realizations. Note that for
both simulation examples, $P_{\text{max}}$ and $\sigma_N^2$ have been chosen
in agreement with commercial IEEE 802.15.4 compliant sensor platforms
\cite{TexasInstruments:07}.
\begin{example}[Arithmetic Mean]\label{ex:accuracy_arith}
  Let $M=25,50,150,250$, the number of nodes $K=M$ and the sensor readings
  uniformly and i.i.d. in
  $\mathcal{X}=[1\,^{\circ}\text{C},30\,^{\circ}\text{C}]\subset\mathcal{S}$. The
  resulting experimental data is depicted in Fig. \ref{fig:accuracy_arith}.
\begin{figure}[t!]
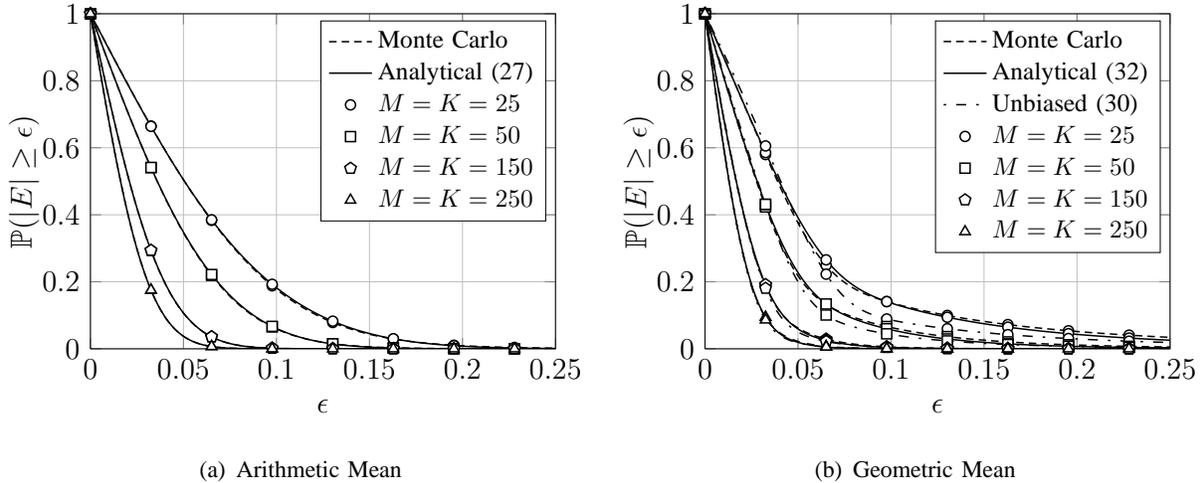

		\centering
		\subfigure[Arithmetic Mean]{\scalebox{0.95}{\input{comac_arith_accuracy.pgf}}\label{fig:accuracy_arith}}
		\subfigure[Geometric Mean]{\scalebox{0.95}{\input{comac_geo_accuracy.pgf}}\label{fig:accuracy_geo}}
		\caption{Monte Carlo evaluation of the outage probabilities ($10\cdot10^3$ realizations) vs. analytical results for different $M=K$.}
		\label{fig:accuracy}
\end{figure}
\end{example}

The plots in Fig. \ref{fig:accuracy_arith} indicate that expression (\ref{eq:prob_approx_gauss}) accurately approximates
the true outage probability $\Prob(|E|\geq\epsilon)$ for all $\epsilon>0$,
since already for relatively short sequence lengths, differences between the
analytical expression and the Monte Carlo simulations are negligible. Furthermore, the plots numerically confirm
the consistency statement of Proposition \ref{prop:arith_consistent}, because the probability curves tend to the ordinate
axis with growing $M$.
\begin{example}[Geometric Mean]\label{ex:accuracy_geo}
  Let
  $\mathcal{S}'\coloneqq[s',s_{\text{max}}]=[0.5\,^{\circ}\text{C},130\,^{\circ}\text{C}]\subset\mathcal{S}$,
  $a=2$,
  $\mathcal{X}=[1\,^{\circ}\text{C},30\,^{\circ}\text{C}]\subset\mathcal{S}'$
  and all other simulation parameters as in Example
  \ref{ex:accuracy_arith}.\footnote{Notice that the corresponding function
    estimation error relies on $\mathcal{S}'$ since desired function
    geometric mean can not be continuously extended onto the entire
    sensing range $\mathcal{S}$.} The resulting experimental data is depicted in
  Fig. \ref{fig:accuracy_geo}.
\end{example}

Similar as for Example \ref{ex:accuracy_arith}, the plots in
Fig. \ref{fig:accuracy_geo} show that (\ref{eq:prob_end}) with
(\ref{eq:prob_sub}) approximates the true outage probability sufficiently accurate, with a negligible deviation for short sequence lengths. In Section \ref{sec:geometric_analysis}, we mentioned that although the geometric mean estimator (\ref{eq:f_estimate}) is applicable in practice, it has the
drawback of only an asymptotic unbiasedness compared to the impractical unbiased
estimator (\ref{eq:f_estimate_un}). Nevertheless, besides a comparison of a
Monte Carlo evaluation of $\Prob(|E|\geq\epsilon)$ using (\ref{eq:f_estimate})
with the analytical result (\ref{eq:prob_sub}), the figure also contains a plot in which (\ref{eq:f_estimate_un}) was used to quantify the drawback.
The difference between (\ref{eq:f_estimate}) and
(\ref{eq:f_estimate_un}) vanishes quickly with increasing $M$, which
confirms Proposition \ref{prop:properties} and Remark \ref{rem:accuracy}. 
%
\begin{remark}\label{rem:length_m}
  Propositions \ref{prop:arith_consistent} and \ref{prop:properties} as well
  as Examples \ref{ex:accuracy_arith} and \ref{ex:accuracy_geo} demonstrate
  that the sequence length $M$ is the crucial design parameter, which
  determines the trade-off between computation accuracy and computation
  throughput.
\end{remark}
%
%
%
\subsection{Comparisons with TDMA}\label{sec:comparison_tdma}
%
%
The numerical examples in the preceding subsection indicate the general
behavior of the proposed analog computation architecture without concrete evidence
regarding the computation performance compared to standard multiple-access
schemes. Therefore, we demonstrate in this subsection the superiority of the proposed
CoMAC architecture by a comparison with an idealized uncoded TDMA scheme. For TDMA, the individual nodes quantize their sensor readings uniformly over $\mathcal{S}$ with
$Q\in\mathds{N}$ $\text{bit}$, followed by binary phase shift keying, such that each sensor has to transmit a bit
stream of length $Q$ to the FC.

To ensure fairness between CoMAC and TDMA, with fixed degrees of freedom
(e.g., bandwidth, symbol duration), both schemes should induce the same costs
per function value computation with respect to transmit energy and
transmit time. 
Therefore, let $T\in\mathds{R}_{++}$ be the common symbol duration and let
$P_{\text{TDMA},k}\in\mathds{R}_{++}$ denote the instantaneous TDMA transmit power on
node $k\in\mathcal{K}$. Then, the transmit times per function value are
$T_{\text{CoMAC}}=MT$ and $T_{\text{TDMA}}=QKT$, whereas the transmit energies
can be written as ${\sf{E}}_{\text{CoMAC},k}=MP_kT$ and
${\sf{E}}_{\text{TDMA},k}=QP_{\text{TDMA},k}T$, respectively. Now, from the fairness conditions $T_{\text{CoMAC}}=T_{\text{TDMA}}$ and ${\sf{E}}_{\text{CoMAC},k}={\sf{E}}_{\text{TDMA},k}$, for all $k\in\mathcal{K}$, it follows $M=QK$ for the CoMAC sequence length and $P_{\text{TDMA},k}=\tfrac{P_kM}{Q}=\tfrac{g_{\varphi}(\varphi_k(X_k))M}{Q}$,
$k\in\mathcal{K}$, for the required instantaneous TDMA transmit powers.

\begin{figure}[t!]
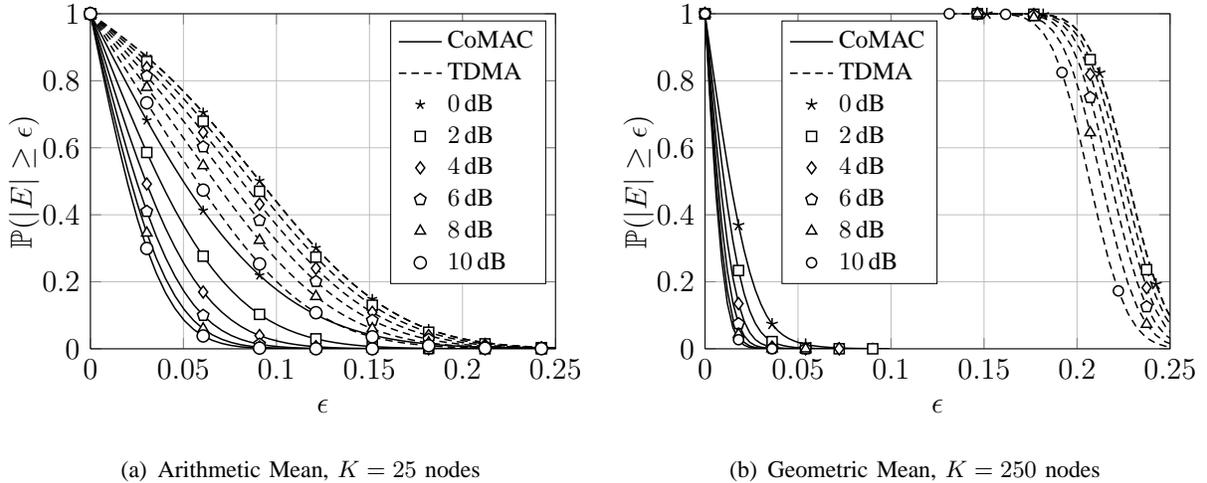

		\centering
		\subfigure[Arithmetic Mean, $K=25$ nodes]{\scalebox{0.95}{\input{comacvstdma_arith_K25.pgf}}\label{fig:small_network_ex}}
		\subfigure[Geometric Mean, $K=250$ nodes]{\scalebox{0.95}{\input{comacvstdma_geo_K250_4.pgf}}\label{fig:medium_network_ex}}
		\caption{CoMAC vs. TDMA: outage probabilities for quantization with $Q=10\,\text{bit}$ (in the case of TDMA), sequence length $M=QK$, and $\SNR_f^{\text{dB}}=0,2,4,6,8,10\,\text{dB}$.}
		\label{fig:network_ex}
\end{figure}

In addition to fairness, requires an adequate comparison the
determination of a common system operating point, which can be done in terms of an average Signal-to-Noise Ratio (SNR). Assume for simplicity that the sensed values $X_k$ are i.i.d. in
$\mathcal{X}$, for all $k$, such that the average received TDMA-SNR per node can be defined as
\begin{equation}
	\SNR_{f}\coloneqq\frac{2M\E\{P_1\}}{\sigma_N^2Q}\;,
	\label{eq:ave_TDMA_SNR}
\end{equation}
which depends on the desired function.
\begin{example}[Small Network Size]\label{ex:comparison1}
  Let $K=25$, $Q=10\,\text{bit}$, the sequence length $M=QK$, and let $P_{\text{max}}$ and $\sigma_N^2$ be chosen such that
  $\SNR_{f}^{\text{dB}}\coloneqq10\log_{10}(\SNR_f)\in\{0,2,4,6,8,10\}$. Furthermore, let the sensor readings be uniformly and
  i.i.d. in $\mathcal{X}=[5\,^{\circ}\text{C},30\,^{\circ}\text{C}]\subset\mathcal{S}$
  and let the desired function be ``arithmetic mean''. The corresponding
  simulation data is depicted in Fig. \ref{fig:small_network_ex}.
\end{example}
\begin{example}[Medium Network Size]\label{ex:comparison2}
  Let $K=250$, the desired function be ``geometric mean'' with
  $\mathcal{S}'=[1\,^{\circ}\text{C},130\,^{\circ}\text{C}]\subset\mathcal{S}$,
  $a=2$ and let all other simulation parameters as in Example
  \ref{ex:comparison2}. The corresponding simulation data is shown in
  Fig. \ref{fig:medium_network_ex}.
\end{example}

Figures \ref{fig:small_network_ex} and \ref{fig:medium_network_ex} indicate
the huge potential of the proposed analog CoMAC scheme for efficiently computing linear and
nonlinear functions over the wireless channel. In both examples, CoMAC entirely
outperforms TDMA with respect to the computation accuracy for different network parameters. It should be clear that the shown performance gains can be replaced by a
computation throughput gain.
\begin{remark}\label{rem:conservative}
  It is important to emphasize that the shown performance gains are quite conservative since the simulated TDMA scheme was idealized
  in many ways. For example, a realistic TDMA would require an established
  protocol stack with considerable amount of overhead per frame (e.g., header,
  synchronization information, check sum) such that the overall TDMA
  transmission time would extend to $T_{\text{TDMA}}=(Q+R)KT$ with a
  certain $R\in\mathds{N}$.
\end{remark} 
%
%
%
%
%
%
%
%
\section{Conclusion} \label{sec:discussion}
In this paper, we proposed a simple analog scheme for efficiently computing functions of the measurements in wireless sensor networks. The main idea of the approach is to exploit the natural superposition property of the wireless channel by letting nodes transmit simultaneously to a fusion center. Applying an appropriate pre-processing function to each sensor reading prior to transmission and a post-processing function to the signal received by the fusion center, which is the superposition of the signals transmitted by the individual nodes, the approach allows the analog computation of a huge set of linear and nonlinear functions over the channel. To relax corresponding synchronization requirements, the nodes transmit some random sequences at a transmit power that is proportional to the respective pre-processed sensor information. As a consequence, only a coarse frame synchronization is required such that the scheme is robust against synchronization errors on the symbol and phase level. The second essential part of the scheme consists of an analog computation-receiver that is designed to appropriately estimate desired function values from the post-processed received sum of transmit energies. Since the estimator has to be matched to the desired function, we considered two canonical function examples and proposed corresponding estimators with good statistical properties.  

Numerical comparisons with a standard TDMA have shown that the proposed analog computation scheme has the potential to achieve huge performance gains in terms of computation accuracy or computation throughput. In addition to the weaker requirements regarding the synchronization of sequences, the scheme needs no explicit protocol structure, which significantly reduces the overhead. Computation schemes following the described design rule are therefore energy and complexity efficient and can be easily
implemented in practice. Finally, the hardware-effort is reduced as well since energy
consuming digital components (e.g., analog-to-digital converters, registers)
are not required. Note that the proposed computation scheme can be used as a building block for more complex in-network processing tasks.
%
%
%
%
%
%
%
%
%
\appendix
%
%
\subsection{Proof of Proposition
  \ref{lem:affine_functions}}\label{app:proof_lem_affine_functions}
%
%
Let $g_{\varphi}'\coloneqq Mg_{\varphi}$ and $h_{\varphi}'\coloneqq 1/M
h_{\varphi}$ so that we have to show that
$h_{\varphi}'(\sum_kg_{\varphi}'(\xi_k))=\sum_k\xi_k$ with
$\xi_k\in[\varphi_{\text{min}},\varphi_{\text{max}}]$,$k\in\mathcal{K}$, holds
if and only if $g_{\varphi}$ and $h_{\varphi}$ are
affine functions. The ``$\Leftarrow$'' direction is
trivial, while the other direction is shown by contradiction. Suppose
$g_{\varphi}'$ is bijective and continuous but not affine. Then there exist
two points $(\xi_1,\dots,\xi_K)$ and $(\tilde{\xi}_1,\dots,\tilde{\xi}_K)$ in $[\varphi_{\text{min}},\varphi_{\text{max}}]^K$
with $\sum_k\xi_k\neq\sum_k\tilde{\xi}_k$ but
$\sum_kg_{\varphi}'(\xi_k)=\sum_kg_{\varphi}'(\tilde{\xi}_k)$. By the last
equation, we have
\begin{equation*}
  \textstyle\sum_k\xi_k=h_{\varphi}'\bigl(\sum_kg_{\varphi}'(\xi_k)\bigr)=h_{\varphi}'\bigl(\sum_kg_{\varphi}'(\tilde{\xi}_k)\bigr)
  =\sum_k\tilde{\xi}_k\;,
\end{equation*}
which however contradicts $\sum_k\xi_k\neq\sum_k\tilde{\xi}_k$. Hence,
$g_{\varphi}'$ is affine and so is $g_{\varphi}$. Moreover, we have
$h_{\varphi}(\sum_kg_{\varphi}'(\xi_k))=h_{\varphi}(Mg_{\varphi}(\sum_k\xi_k)+\tilde{c})$
for some $\tilde{c}\in\mathds{R}$, from which we conclude that $h_{\varphi}$
is an affine function as well with $h_{\varphi}\equiv g_{\varphi}^{-1}-c$ and
some constant $c\in\mathds{R}$ that depends on $g_{\varphi}$.
%
%
%
%
\subsection{Proof of Lemma \ref{lem:beta}}\label{app:proof_lem_beta}
%
%
Since $\Delta_3\sim\chi^2_{2M}$, the probability density of $\Delta_3$ is
$p_{\Delta_3}(x)
=\frac{1}{\sigma_N^{2M}\,\Gamma(M)}x^{M-1}\e^{-x/\sigma_N^2}\mathds{1}_{[0,\infty)}(x)$,
where $\Gamma(z)$ with $\Real\{z\}>0$ is used to denote the Gamma
function. Hence, one obtains
\begin{equation}
  \E\bigl\{\psi\bigl({\textstyle{\frac{\Delta_3}{\alpha_{\text{geo}}M}}}\bigr)\bigr\}
=\tfrac{1}{\sigma_N^{2M}\,\Gamma(M)}\int_0^{\infty}\!x^{M-1}\exp\Bigl(-\Bigl(\tfrac{\alpha_{\text{geo}}KM-\sigma_N^2\log_{\e}(a)}{\sigma_N^2\alpha_{\text{geo}}KM}\Bigr)x\Bigr)\,dx.
    \label{eq:expec1}
\end{equation}
Now assume $\sigma_N^2\log_{\e}(a)<\alpha_{\text{geo}} KM$ and note that
$\Gamma(z) = \int_0^{\infty}x^{z-1}\e^{-x}dx =
k^z\int_0^{\infty}x^{z-1}\e^{-kx}dx$, $\Real\{z\}>0$, which holds for any
$\Real\{k\}>0$ \cite{Abramowitz:Stegun:72}. So substituting this into
(\ref{eq:expec1}) with an appropriately chosen $k$ proves (i). As for (ii), if
$\sigma_N^2\log_{\e}(a)<\alpha_{\text{geo}} KM$, then it follows from (i) that
$\lim_{M\to\infty}\left(\tfrac{\alpha_{\text{geo}} KM}{\alpha_{\text{geo}}
    KM-\sigma_N^2\log_{\e}(a)}\right)^M=\lim_{M\to\infty}(1+\frac{u}{M})^{-M}=\e^{-u}$, 
where $u\coloneqq -\tfrac{\sigma_N^2\log_{\e}(a)}{\alpha_{\text{geo}}K}$.
%
%
%
%
%
\subsection{Proof of Lemma \ref{lem:Xi_phi}}\label{app:proof_lem_log1}
%
%
Let $\mathcal{X}$ be an arbitrary compact set and $K<\infty$ any fixed natural
number. By Section \ref{sec:computableWMAC} and Definition
\ref{def:geometric_estimator}, we know that
$\Xi|\ve{x}=\psi(\frac{\Delta|\ve{x}}{\alpha_{\text{geo}}M})=a^{\frac{1}{\alpha_{\text{geo}}
    KM}\Delta|\ve{x}}$, $K,\alpha_{\text{geo}}>0$, $a>1$. Since $\psi$ is
continuous and strictly increasing,
$P_{\Xi}(\xi|\ve{x})=\Prob(\Xi\leq\xi|\ve{X}=\ve{x})=\Prob(\Delta\leq\alpha_{\text{geo}}
KM\log_a({\xi})|\ve{X}=\ve{x})=P_{\Delta}(\alpha_{\text{geo}}
KM\log_a(\xi)|\ve{x})$, $\xi>0$. Thus, bearing in mind Corollary
\ref{cor:degenerated}, we can conclude that, as $M$ sufficiently large, $\Delta|\ve{x}$
can be approximated by a random variable
$\tilde{\Delta}|\ve{x}\sim\mathcal{N}_{\mathds{R}}(M\sigma_N^2,\sigma_{\Delta|\ve{x}}^2)$. An
immediate consequence of this is that for sufficiently large values of $M$,
the distribution function of $\Delta|\ve{x}$ can be approximated by
$P_{\tilde{\Delta}}(\delta|\ve{x})
=\frac{1}{2}+\frac{1}{2}\erf\bigl(\frac{\delta-M\sigma_N^2}{\sqrt{2}\sigma_{\Delta|\ve{x}}}\bigr)$
(i.e., $P_{\Delta}(\delta|\ve{x})\approx
P_{\tilde{\Delta}}(\delta|\ve{x})$). Moreover, for $M$ large enough, the
Mann-Wald theorem \cite[p.\,356]{Shiryaev:96} implies
$P_{\Xi}(\xi|\ve{x})\approx P_{\tilde{\Xi}}(\xi|\ve{x})=
P_{\tilde{\Delta}}(\alpha_{\text{geo}}
KM\log_a(\xi)|\ve{x})$, where ($\xi\in\mathds{R}_{++}$)
\begin{equation}
    P_{\tilde{\Delta}}(\alpha_{\text{geo}} KM\log_a(\xi)|\ve{x})=\frac{1}{2}+\frac{1}{2}\erf\left(\frac{\alpha_{\text{geo}}KM\log_{\e}(\xi)-\sigma_N^2M\log_{\e}(a)}{\sqrt{2}\,\log_{\e}(a)\sigma_{\Delta|\ve{x}}}\right)\,.
    \label{eq:dist_approx_Xi}
\end{equation}
Note that (\ref{eq:dist_approx_Xi}) describes the distribution function of a
log-normally distributed random variable with parameters
$\frac{\sigma_N^2\log_{\e}(a)}{\alpha_{\text{geo}} K}\eqqcolon\mu_{\Xi}$ and
$\bigl(\frac{\log_{\e}(a)}{\alpha_{\text{geo}}KM}\sigma_{\Delta|\ve{x}}\bigr)^2\eqqcolon\sigma_{\Xi|\ve{x}}^2$.
Thus $\Xi|\ve{x}$ is approximated by a random variable
$\tilde{\Xi}|\ve{x}\sim\mathcal{LN}(\mu_{\Xi},\sigma_{\Xi|\ve{x}}^2)$.
%
%
%
%
%
\subsection{Proof of Proposition \ref{prop:prob_abs_cond}}\label{app:proof_prop_prob}
%
%
Note that it is sufficient to show (\ref{eq:prob_sub}). Because
$|E|\ve{x}|=|\gamma(\ve{x})^{-1}\Xi|\ve{x}-\beta(\ve{x})|$ is continuous in
$\Xi|\ve{x}$, Lemma \ref{lem:Xi_phi} and the Mann-Wald theorem allow for the
approximation of $|\Xi|\ve{x}|$ by
$|\tilde{E}|\ve{x}|=|\gamma(\ve{x})^{-1}\tilde{\Xi}|\ve{x}-\beta(\ve{x})|$,
where the probability distribution function of
$\tilde{\Xi}|\ve{x}\sim\mathcal{LN}(\mu_{\Xi},\sigma_{\Xi|\ve{x}}^2)$ is given
by (\ref{eq:dist_approx_Xi}). Since $0<\beta(\ve{x}),\gamma(\ve{x})<\infty$,
we have
$\Prob(|E|\geq\epsilon|\ve{X}=\ve{x})\approx\Prob(|\tilde{E}|\geq\epsilon|\ve{X}=\ve{x})=1-\Prob(-\epsilon<\tilde{E}<\epsilon|\ve{X}=\ve{x})=1-\Prob(-\epsilon<\gamma(\ve{x})^{-1}\tilde{\Xi}-\beta(\ve{x})<\epsilon|\ve{X}=\ve{x})$
which leads to
\begin{equation}
  \Prob(|\tilde{E}|\geq\epsilon|\ve{X}=\ve{x})=
  \begin{cases}
    1-P_{\tilde{\Xi}}\bigl(\rho^+(\ve{x},\epsilon)|\ve{x}\bigr)+P_{\tilde{\Xi}}\bigl(\rho^-(\ve{x},\epsilon)|\ve{x}\bigr),&0<\epsilon<\beta(\ve{x})\\
    1-P_{\tilde{\Xi}}\bigl(\rho^+(\ve{x},\epsilon)|\ve{x}\bigr),&\beta(\ve{x})\leq\epsilon<\infty
  \end{cases}
  \label{eq:cases}
\end{equation}
where $\rho^+(\ve{x},\epsilon)\coloneqq\gamma(\ve{x})(\beta(\ve{x})+\epsilon)$
and
$\rho^-(\ve{x},\epsilon)\coloneqq\gamma(\ve{x})(\beta(\ve{x})-\epsilon)$. Inserting
the right-hand side of (\ref{eq:dist_approx_Xi}) into expression
(\ref{eq:cases}) and using $\erfc(x)=1-\erf(x)$, for all $x\in\mathds{R}$,
shows (\ref{eq:prob_sub}) and thus completes the proof.
%
%
%
%
%
%
%
%
%
%
%
\bibliographystyle{IEEEtran}
\bibliography{IEEEabrv,mario_goldenbaum}
\end{document}